\documentclass[]{aa}

\usepackage{graphicx}
\usepackage{fixltx2e}

\usepackage{txfonts}
\usepackage[]{hyperref}
\begin{document} 

   \title{Inflowing gas onto a compact obscured nucleus in Arp 299A\thanks{\emph{Herschel} is an ESA space observatory with science instruments provided by European-led Principal Investigator consortia and with important participation from NASA.}}
   \subtitle{Herschel spectroscopic studies of H$_{2}$O and OH}
   \author{N. Falstad \inst{1}
          \and
          E. Gonz\'alez-Alfonso \inst{2}
          \and
          S. Aalto \inst{1}
          \and
          J.~Fischer \inst{3}
\fnmsep
          }

   \institute{Department of Earth and Space Sciences, Chalmers University of Technology, Onsala Space Observatory,
     439 92 Onsala, Sweden \\
     \email{niklas.falstad@chalmers.se}
     \and
     Universidad de Alcal\'a de Henares,
     Departamento de F\'{\i}sica, Campus Universitario, E-28871 Alcal\'a de
     Henares, Madrid, Spain
     \and
     Naval Research Laboratory, Remote Sensing Division, 4555
     Overlook Ave SW, Washington, DC 20375, USA
   }

   \date{}

 
  \abstract
    {}
   {We probe the physical conditions in the core of Arp~299A and try to put constraints to the nature of its nuclear power source.}
   {We used \emph{Herschel} Space Observatory far-infrared and submillimeter observations of H$_{2}$O and OH rotational lines in Arp~299A to create a multi-component model of the galaxy. In doing this, we employed a spherically symmetric radiative transfer code.}
   {Nine H$_{2}$O lines in absorption and eight in emission as well as four OH doublets in absorption and one in emission, are detected in Arp~299A. No lines of the $^{18}$O isotopologues, which have been seen in compact obscured nuclei of other galaxies, are detected. The absorption in the ground state OH $^{2}\Pi_{3/2}-{^{2}\Pi}_{3/2}\, \frac{5}{2}^+-\frac{3}{2}^-$ doublet at $119$~$\mu$m is found redshifted by ${\sim}175$~km\,s$^{-1}$ compared with other OH and H$_{2}$O lines, suggesting a low excitation inflow. We find that at least two components are required in order to account for the excited molecular line spectrum. The inner component has a radius of $20-25$~pc, a very high infrared surface brightness ($\gtrsim3\times10^{13}$~L$_{\sun}~$kpc$^{-2}$), warm dust ($T_{\mathrm{d}}>90$~K), and a large H$_{2}$ column density (N$_{\mathrm{H_{2}}}>10^{24}$~cm$^{-2}$). The modeling also indicates high nuclear H$_{2}$O ($1-5 \times 10^{-6}$) and OH ($0.5-5\times 10^{-5}$) abundances relative to H nuclei. The outer component is larger ($50-100$~pc) with slightly cooler dust ($70-90$~K) and molecular abundances that are about one order of magnitude lower. In addition to the two components that account for the excited OH and H$_{2}$O lines, we require a much more extended inflowing component to account for the  OH $^{2}\Pi_{3/2}-{^{2}\Pi}_{3/2}\, \frac{5}{2}^+-\frac{3}{2}^-$ doublet at $119$~$\mu$m.}
   {The Compton-thick nature of the core makes it difficult to determine the nature of the buried power source, but the high surface brightness indicates that it is either an active galactic nucleus and/or a dense nuclear starburst. Our results are consistent with a composite source. The high OH/H$_{2}$O ratio in the nucleus indicates that ion-neutral chemistry induced by X-rays or cosmic-rays is important. Finally we find a lower limit to the $^{16}$O/$^{18}$O ratio of $400$ in the nuclear region, possibly indicating that the nuclear starburst is in an early evolutionary stage, or that it is fed through a molecular inflow of, at most, solar metallicity.}


   \maketitle
%

\section{Introduction}
Luminous infrared galaxies \citep[LIRGs; see the review by][]{san96} are galaxies with infrared (IR) luminosities in excess of $10^{11}$~L$_{\odot}$, powered by active galactic nuclei (AGN) or circumnuclear starbursts. An emerging subclass of LIRGs host compact obscured nuclei (CONs) where luminosities in excess of $10^{9}$~L$_{\odot}$ originate in compact ($d<100$~pc) dust obscured cores \citetext{e.g. \citealp{aal12,cos10}; \citealp[][hereafter G-A12]{gon12}; \citealp{fal15}}\defcitealias{gon12}{G-A12}. The central regions of these objects are heated by a process in which the radiation is absorbed and re-emitted at progressively longer wavelengths until the dust is optically thin to its own radiation \citetext{\citealp{rol11}; \citetalias{gon12}}. Due to this radiative trapping, the power source is hidden from direct observations over a broad range of wavelengths, making it hard to discern whether it is star formation or AGN activity. Determining the nature of these power sources could potentially aid our understanding of galaxy evolution greatly. 

One method to identify, characterize, and indirectly study the cores of CONs is to use molecules that couple well to the IR radiation field, for example water (H$_{2}$O) and hydroxyl (OH), to probe the warm dust in these regions \citetext{e.g. \citealp{gon04,gon08}; \citetalias{gon12}; \citealp{fal15}}. There are indeed indications that sources with a strong OH $65$~$\mu$m $^{2}\Pi_{3/2}-{^{2}\Pi}_{3/2}\, \frac{9}{2}-\frac{7}{2}$ doublet represent the most buried stage of starburst-AGN co-evolution \citep{gon15}. Furthermore, submillimeter (submm) H$_{2}$O lines have been detected in numerous nearby \citep[e.g.][]{yan13} and high-redshift galaxies \citep[e.g.][]{imp08,omo11,omo13,bra11,van11}. In some cases, for example in Mrk~231 \citep{van10,gon10} and Zw~049.057 \citep{fal15}, these H$_{2}$O lines have fluxes comparable to those of the carbon monoxide (CO) lines.

\subsection{The interacting system Arp 299}
Arp~299 is an interacting system consisting of the two galaxies IC~694 and NGC~3690, $22$\arcsec\ apart, whose nuclei (Arp~299A and Arp~299B) as well as an overlap region (Arp~299C) are strong near-IR and radio emitters due to intense star formation \citep{geh83,tel85}. Based on \ion{H}{i} observations, \citet{nor97} determined the systemic velocity of the system to be $3121$~km\,s$^{-1}$, yielding a redshift of $z=0.010411$. At a distance of $44.8$~Mpc (linear scale: $217$~pc/arcsec) \citep[][assuming $H_{0}=73$~km\,s$^{-1}$\,Mpc$^{-1}$]{fix96,nor97} the IR luminosity of the system is $L_{\mathrm{IR}}\approx 6.6 \times 10^{11}$~L$_{\odot}$ \citep{san03} of which Arp~299A contributes about $40-50\%$ \citep{alo00,cha02}. All three regions also exhibit bright molecular emission from hydrogen cyanide (HCN) $J=1-0$ and carbon monoxide (CO) \citep{sar91,aal97}. The molecular gas in the compact nucleus of Arp~299A is suggested by \citet{aal97} to be unusually warm and dense. The system also exhibits hydroxyl (OH) megamaser activity \citep{baa85} from a rotating disk of size $\lesssim1$\arcsec\ in Arp~299A \citep{baa90}. A water (H$_{2}$O) megamaser in the system was reported by \citet{hen05} and interferometric observations by \citet{tar07,tar11} showed that the emission mainly originates in Arp~299B with a second hotspot in the inner regions of Arp~299A. Using CO and HCN as probes, \citet{ros14} studied molecular gas heating in the Arp~299 system and found that Arp~299A contains more warm gas than the other two regions. They further concluded that the excitation in Arp~299B and Arp~299C is consistent with heating from ultraviolet photons in photon-dominated regions (PDRs) while the excitation in Arp~299A requires an additional heating mechanism with mechanical heating being the most likely.

Based on optical, radio, and IR data, \citet{geh83} found that the emission from Arp~299, except for the radio source in Arp~299A, is consistent with starbursts at different locations. A compact condensation of molecular gas in the nucleus of Arp~299A was suggested by \citet{sar91} to harbor an AGN. In their X-ray measurements of the system, \citet{del02} found evidence for a Compton-thick AGN, but with the spatial resolution offered by \emph{BeppoSAX} they could not determine its location in the system. Higher spatial resolution observations with \emph{Chandra} and \emph{XMM-Newton} reveal that Arp~299B is the likely home of this AGN \citep{zez03,bal04}. \citet{bal04}, however, suggested that Arp~299A might also harbor an AGN and based on milliarcsecond radio observations \citet{per10} concluded that there is in fact a buried low-luminosity AGN in Arp~299A. Mid-infrared observations of the system are consistent with deeply embedded star formation in the nucleus of Arp~299A and an AGN surrounded by regions of star formation in Arp~299B \citep{gal04,alo09}. Subsequent high spatial resolution mid-IR observations by \citet{alo13} also show evidence of AGN activity in both nuclei and they estimate that the AGN in Arp~299A is five times less luminous than the one in Arp~299B, this corresponds to ${\sim}5\%$ of the total luminosity of Arp~299A. In their \emph{NuSTAR} observations \citet{pta15} found that the hard X-ray emission ($E>10$~keV) of the system is dominated by Arp~299B, with no significant emission from the position of Arp~299A. Their interpretation is that an AGN in Arp~299A must either be heavily obscured, with $N_{\mathrm{H}}>10^{24}$~cm$^{-2}$, or have a much lower luminosity than the AGN in Arp~299B.

In this paper, we model the dusty core of the potential CON in Arp~299A using spectroscopic observations taken with the Photodetector Array Camera and Spectrometer \citep[PACS;][]{pog10} and the Spectral and Photometric Imaging Receiver \citep[SPIRE;][]{gri10} on the \emph{Herschel} Space Observatory \citep{pil10}. The observations are described in Sect. \ref{sec:observations}, and our models are described in Sect. \ref{sec:models}. The model results are discussed in Sect. \ref{sec:discussion}, and our main conclusions are presented in Sect. \ref{sec:conclusions}.

\section{Observations and results}\label{sec:observations}
\begin{figure}  
  \centering
  \includegraphics[trim={2cm 0.5cm 3.5cm 2cm},clip, width=8.0cm]{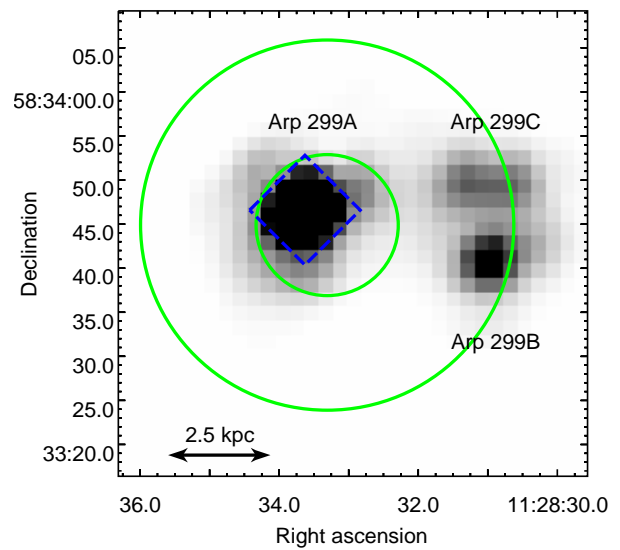}
  \caption{SPIRE beams and the field of view of the central PACS spaxel overlaid on an archival PACS $70$~$\mu$m image of Arp~299. The large circle represents the largest SPIRE beam with a full-width-half-maximum (FWHM) of ${\sim}42$\arcsec\ at a wavelength of ${\sim}670$~$\mu$m and the smaller circle represents the smallest beam with a FWHM of ${\sim}16$\arcsec\ at a wavelength of ${\sim}200$~$\mu$m. The dashed square represents the central $9.4$\arcsec\ spaxel of the PACS spaxel array. Evidently the far-IR emission of the system is dominated by Arp~299A, which is unresolved by PACS at $70$~$\mu$m. The coordinates are in the J2000.0 system.}
  \label{fig:spirebeam}
\end{figure}

The observations of Arp~299A were taken using the PACS and SPIRE instruments on the \emph{Herschel} Space Observatory. Most of the PACS observations were conducted on 2012 October 31 as part of the Hermolirg OT2 project (PI: E. Gonz\'alez-Alfonso). The OH $119$~$\mu$m $^{2}\Pi_{3/2}-{^{2}\Pi}_{3/2}\, \frac{5}{2}-\frac{3}{2}$ and H$_{2}$O $187$~$\mu$m $4_{13}\!\rightarrow\!4_{04}$ transitions were observed on 2011 November 21 as part of the OT1 projects OT1\_shaileyd\_1 (PI: S. Hailey-Dunsheath) and OT1\_rmeijeri\_1 (PI: R. Meijerink), respectively. All observations were performed in high spectral sampling, range spectroscopy mode. The SPIRE observation, already reported by \citet{ros14}, was conducted on 2010 June 27 as part of the OT key program Hercules (PI: P.P. van der Werf) with a single pointing centered on Arp~299A. The observation was taken in high spectral resolution, sparse image sampling mode with a resolution of $1.2$~GHz in both observing bands ($447-989$~GHz and $958-1545$~GHz). A total of 35 repetitions (70 FTS scans) were performed, resulting in a total on-source integration time of $4662$~s. The SPIRE beams and the field of view of the central PACS spaxel, overlaid on a PACS $70$~$\mu$m archival image, are shown in Fig. \ref{fig:spirebeam}. The PACS observations were well pointed while the SPIRE observation was mispointed by ${\sim}3$\arcsec. The flux loss in the short wavelength array, the detector with the smallest beam, due to this mispointing should be less than $10$\% \citep{val14}.

\begin{figure}  
  \centering
  \includegraphics[width=8.0cm]{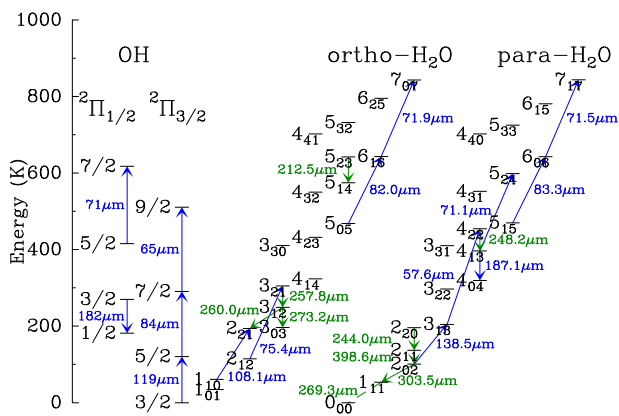}
  \caption{Energy diagram for H$_{2}$O and OH. Transitions detected with PACS and SPIRE are indicated with solid blue and green arrows, respectively. The undetected H$_{2}$O $269$~$\mu$m $1_{11}\!\rightarrow\!0_{00}$ transition is indicated with a dashed green line. Upward and downward arrows indicate absorption and emission lines, respectively.}
  \label{fig:energy_diagram}
\end{figure}

\begin{figure}  
  \centering
  \includegraphics[width=8.0cm]{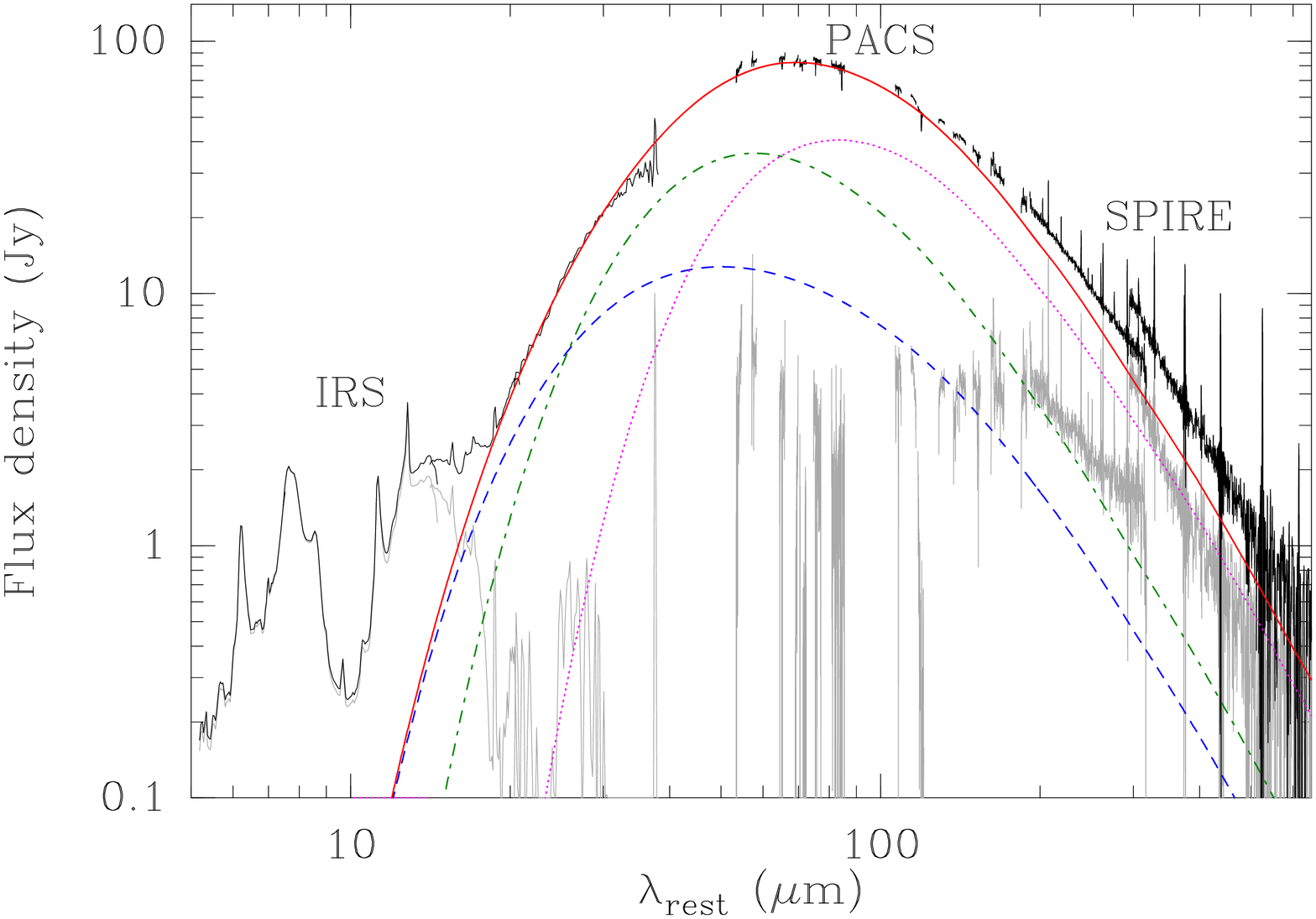}
  \caption{Spectral energy distribution of Arp~299A from mid-IR to submillimeter wavelengths. Data from PACS and SPIRE on \emph{Herschel}, and
IRS on \emph{Spitzer} are shown. The models discussed in Sect. \ref{sec:models} are included where the dashed blue and dashed-dotted green curves represent the inner and outer components, respectively. The dotted magenta curve represents the extended component and the solid red curve is the sum of all model components. Gray curves are the residual spectra after model subtraction.}
  \label{fig:continuum}
\end{figure}

A summary of all observations, with observation identification numbers (IDs), total durations, and the observed wavelength ranges, is provided in Table \ref{tab:obslog}. All detected transitions are indicated in the energy diagram in Fig. \ref{fig:energy_diagram}. A combination of all spectral ranges observed with PACS and SPIRE are presented in Fig. \ref{fig:continuum} together with archival data from the Infrared Spectrograph (IRS) on \emph{Spitzer} \citep{alo09} and the modeled dust continuum discussed in Sect. \ref{sec:models}. Spectroscopic parameters for line identification and radiative transfer modeling were obtained from the JPL \citep{pic98} and CDMS \citep{mul01,mul05} catalogs.

\begin{table*} 
  \caption{Summary of Herschel observations.}             
  \label{tab:obslog}      
  \centering                          
  \begin{tabular}{l c c c c}        
    \hline\hline                 
    Instrument & Observation ID & Date & Duration & Wavelength ranges (rest) \\  
    & & YYYY-MM-DD & (s) & ($\mu$m) \\
    \hline                        
    PACS  & 1342232605 & 2011-11-21 & 976 & $117.6-120.7$ \\ 
    PACS  & 1342232607 & 2011-11-21 & 4759 & $184.3-188.6$ \\ 
    PACS  & 1342254239 & 2012-10-31 & 841 & $53.4-54.8, 68.7-70.0$ \\
    & & & & $106.7-109.5, 137.3-139.9$ \\ 
    PACS  & 1342254240 & 2012-10-31 & 3028 & $80.9-82.5, 82.3-84.2, 83.8-85.5$ \\
    & & & & $161.7-165.0, 164.7-168.4, 167.6-171.0$ \\
    PACS  & 1342254241 & 2012-10-31 & 1151 & $57.1-58.4, 64.5-66.1$ \\
    & & & & $114.3-116.8, 128.9-132.1$ \\
    PACS  & 1342254242 & 2012-10-31 & 2198 & $70.4-72.4, 74.8-75.9, 75.8-77.2$ \\
    & & & & $140.8-144.9, 149.5-151.7, 151.5-154.4$ \\
    SPIRE & 1342199248 & 2010-06-27 & 4964 & $192-310, 300-664$ \\ 
    \hline                                   
  \end{tabular}
\end{table*}

\subsection{Data reduction}\label{sec:datareduction}
\emph{Observations with PACS:} the data reduction was done with the \emph{Herschel} interactive processing environment \citep[HIPE;][]{ott10} version 13.0.0 using the standard background normalization pipeline for chopped line scans and short range scans. At a distance of $44.8$~Mpc, the nuclear far-IR emission in Arp~299A is spatially unresolved in the central $9.4$\arcsec\ (${\sim}2$~kpc) spaxel of the PACS $5$~x~$5$ spaxel array. Because the central spaxel is smaller than the point spread function of the spectrometer, the central spectrum was extracted using a point source correction task available in HIPE version 13.0.0 in order to obtain the full flux. In the OH $^{2}\Pi_{3/2}-{^{2}\Pi}_{3/2}\, \frac{5}{2}-\frac{3}{2}$ doublet at $119$~$\mu$m, line absorption and emission is seen outside of the central spaxel, indicating the presence of a spatially more extended, low-excitation molecular component. For this spectral range we have used the sum of the central nine spaxels in order to obtain all the line flux, but this might also introduce contaminating flux from Arp~299B. We finally fitted the continuum and lines in each spectral range with polynomial baselines and Gaussian curves, respectively. Line fluxes, measured and inferred intrinsic Gaussian line widths, and continuum levels for the PACS observations are listed in Table \ref{tab:pacslines}. Upper limits on the $^{18}$OH and H$_{2}^{18}$O line fluxes were calculated using the instrument resolution and assuming a Gaussian profile. The $3\sigma$ peak flux density limits and flux limits for these lines are presented in Table \ref{tab:18oh}.

\emph{Observations with SPIRE:} the data reduction was done with the standard single pointing pipeline in HIPE 13.0.0. There is a noticeable mismatch in the continuum level in the overlap region between the two SPIRE detectors (see Fig. \ref{fig:continuum}). This is due to a discontinuity in the wavelength dependent beam size between the detectors, and indicates that there is flux outside of the beam of the short wavelength array. This might mean that the source is actually partially extended, or that the detector with the larger beam size picks up flux from the other components of the system (see Fig. \ref{fig:spirebeam}). We expect that the H$_{2}$O emission lines that are of interest here arise in a relatively compact region around the nucleus of Arp~299A (though more extended than the absorption lines, see Sect. \ref{sec:models}), but line emission from Arp~299B/C is also possible. The flux in the H$_{2}$O $303$~$\mu$m  $2_{02}\!\rightarrow\!1_{11}$ line, which is located in the overlap region of SPIRE, shows ${\sim}30\%$ more flux in the long wavelength array than in the short wavelength array. To avoid contamination from Arp~299B/C we have used the flux from the short wavelength array, which has the smaller beam size. The full-width-half-maximum of the beam at the wavelength of the H$_{2}$O $399$~$\mu$m $2_{11}\!\rightarrow\!2_{02}$ line, which is only detected by the long wavelength array, is ${\sim}40$\arcsec, and it is likely contaminated by flux from the other components of the system.

We extracted line fluxes using FTFitter\footnote{https://www.uleth.ca/phy/naylor/index.php?page=ftfitter}, an IDL based linefitter for Fourier transform spectrometer (FTS) spectra. A polynomial baseline was subtracted from the spectrum of each detector and the spectral lines were then simultaneously fitted using Gaussian profiles convolved with the FTS instrumental line shape (a sinc function). Unresolved lines were fitted using the instrumental line shape only. Line fluxes and continuum levels for the H$_{2}$O lines observed with SPIRE are listed in Table \ref{tab:spirelines}. Upper limits on the H$_{2}^{18}$O line fluxes in the SPIRE range were calculated using a sinc function assuming the instrumental resolution. The $3\sigma$ peak flux density limits and flux limits for those H$_{2}^{18}$O transitions that are well separated from other strong lines are presented in Table \ref{tab:18oh}.

\subsection{H$_{2}$O}
\begin{figure*}
  \centering
  \includegraphics[width=18.0cm]{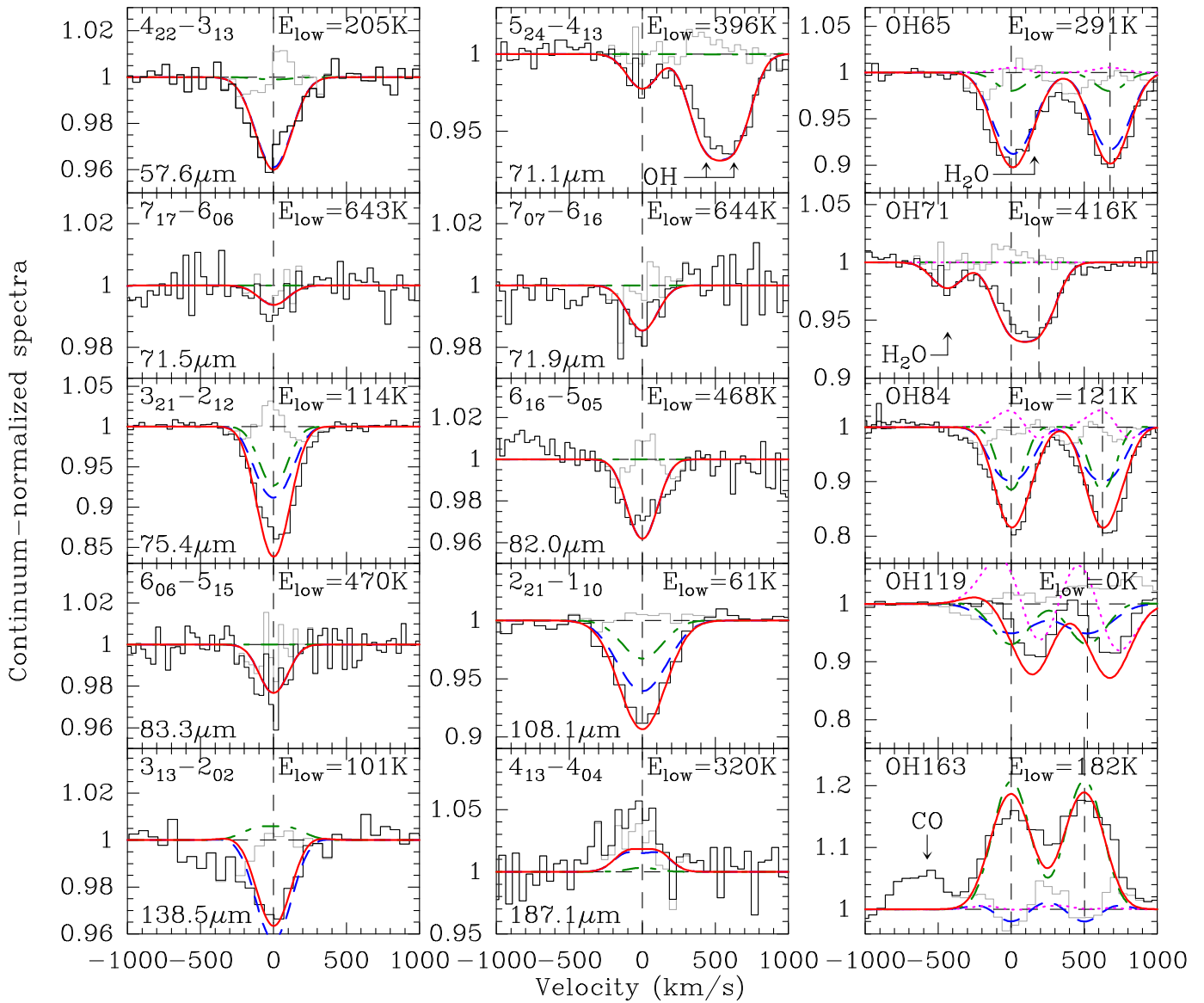}
  \caption{H$_{2}$O and OH lines observed in Arp~299A with PACS. The black histograms are the observed, continuum-normalized, spectra. The velocity scale for the OH lines refers the blue component in each doublet. The best fit models from Sect. \ref{sec:models} are also included. The dashed blue and dashed-dotted green curves denote the models for the inner and outer components, respectively. The dotted magenta curve in the OH spectra represents the tentative extended component and the solid red curve denotes the sum of the models. Gray histograms are the residual spectra after model subtraction.}
  \label{fig:h2o_oh}
\end{figure*}

\begin{figure}  
  \centering
  \includegraphics[width=8.0cm]{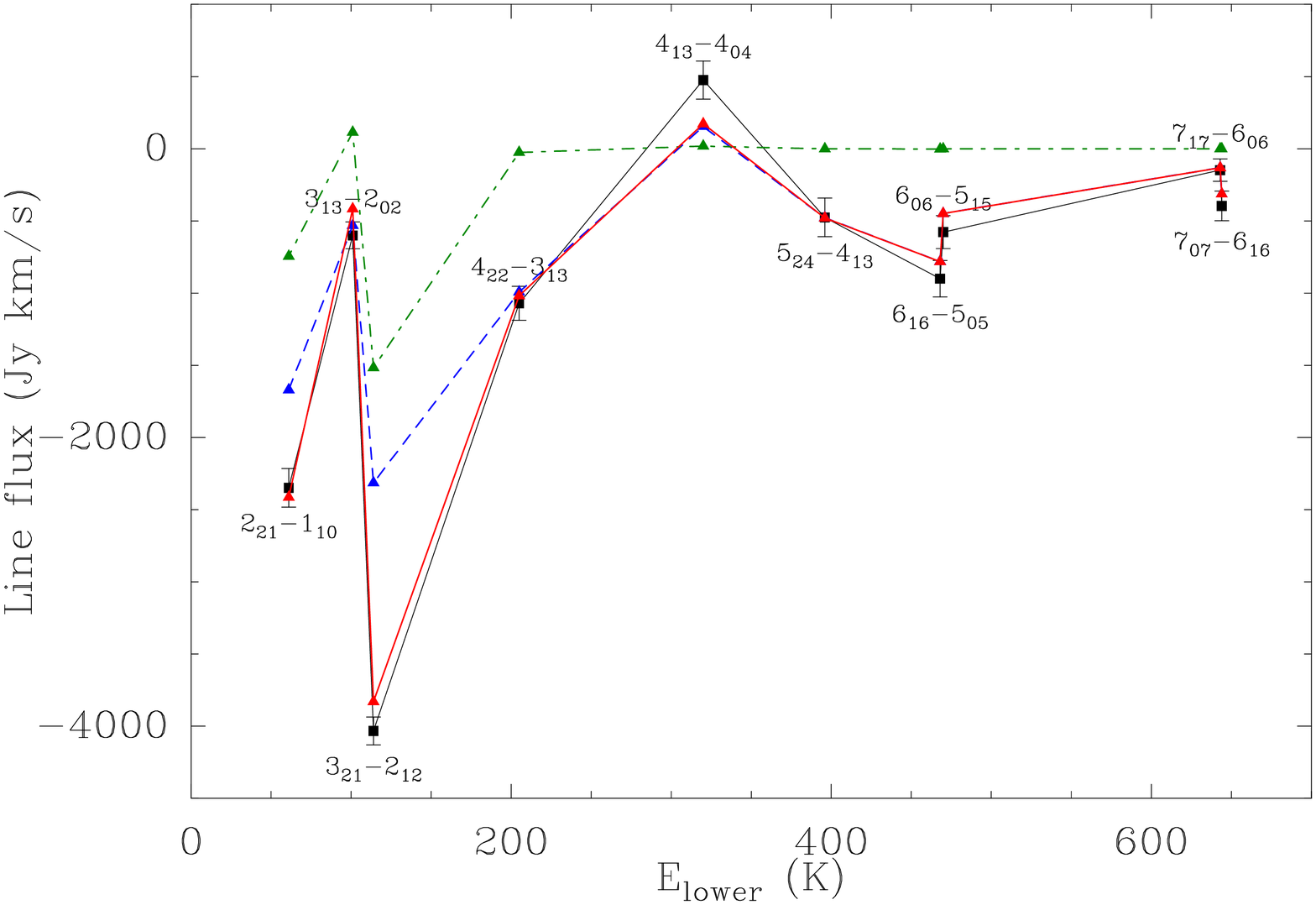}
  \caption{Spectral line energy distribution of the H$_{2}$O lines detected with PACS, the black squares represent the data. Model predictions are included with the dashed blue and dashed-dotted green curves denoting the models for the inner and outer components, respectively. The solid red curve is the sum of these two models.}
  \label{fig:h2o_sled_pacs}
\end{figure}

Nine H$_{2}$O absorption lines and one emission line were detected with PACS. The single emission line lies at $187$~$\mu$m, the longest wavelength of all H$_{2}$O lines detected with PACS. The absorptions have lower (pre-transition) energy levels up to ${\sim} 650$~K while the emission line has an upper energy level of ${\sim}400$~K. The spectral ranges around the lines are shown in Fig. \ref{fig:h2o_oh}, the spectral line energy distribution is presented in Fig. \ref{fig:h2o_sled_pacs}, and the measured line parameters are summarized in Table \ref{tab:pacslines}. There is no systematic velocity shift among the  H$_{2}$O lines and within $2\sigma$ all but one of the lines are consistent with the galaxy redshift of $z=0.010411$ \citep{nor97}. The H$_{2}$O $139$~$\mu$m $3_{13}\!\rightarrow\!2_{02}$ line has a possible blushifted line wing extending down to ${\sim}-600$~km\,s$^{-1}$, but this spectral range is noisy and the feature is at the $1\sigma$-level. We see no obvious contamination by other species in any of the lines.

\begin{figure*}
  \centering
  \includegraphics[width=18.0cm]{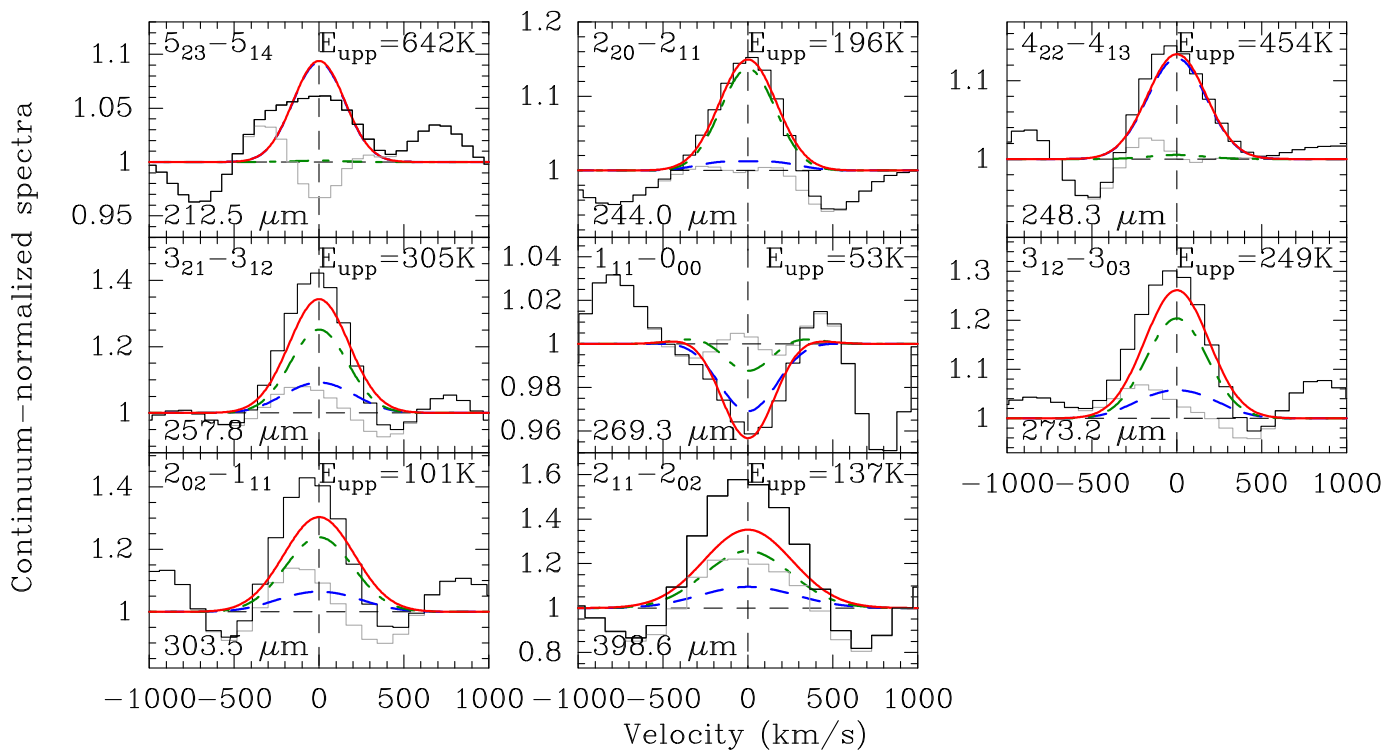}
  \caption{H$_{2}$O lines observed in Arp~299A with SPIRE. The black histograms are the observed, continuum-normalized, spectra. The best fit models from Sect. \ref{sec:models} are also included. The dashed blue and dashed-dotted green curves denote the models for the inner and outer components, respectively, and the solid red curve denotes the sum of the models. Gray histograms are the residual spectra after model subtraction.}
  \label{fig:h2o_spire}
\end{figure*}

\begin{figure}  
  \centering
  \includegraphics[width=8.0cm]{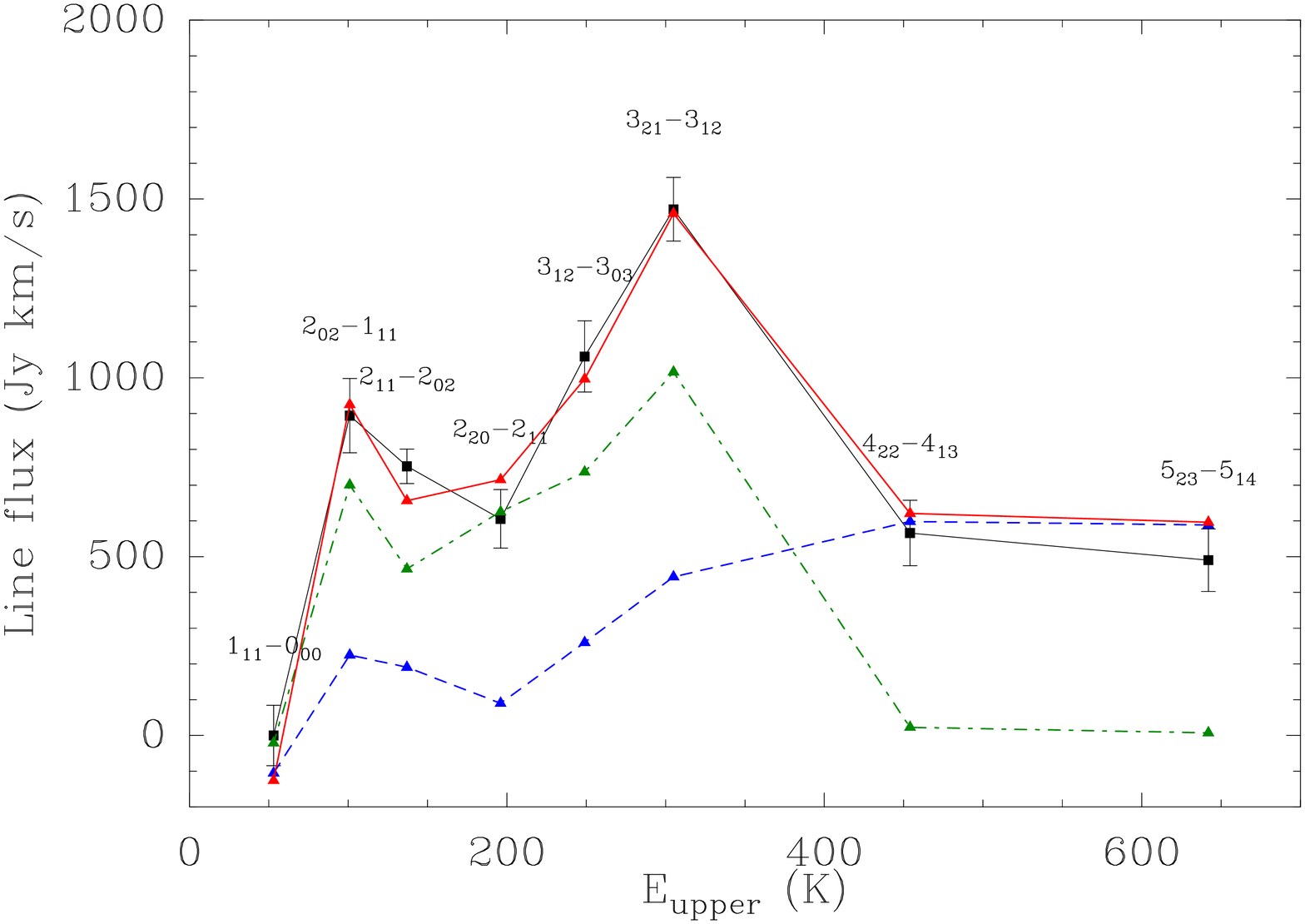}
  \caption{Spectral line energy distribution of the H$_{2}$O lines detected with SPIRE, the black squares represent the data. Model predictions are included with the dashed blue and dashed-dotted green curves denoting the models for the inner and outer components, respectively. The solid red curve is the sum of these two models.}
  \label{fig:h2o_sled}
\end{figure}

Seven H$_{2}$O transitions, with upper energy levels up to ${\sim} 650$~K, were detected in emission with SPIRE. The line fluxes are summarized in Table \ref{tab:spirelines}, the spectral ranges around the lines are presented in Fig. \ref{fig:h2o_spire}, and the spectral line energy distribution is shown in Fig. \ref{fig:h2o_sled}. One of the transitions, H$_{2}$O $399$~$\mu$m $2_{11}\!\rightarrow\!2_{02}$, was detected with the spectrometer long wavelength array; due to the larger beam size for this detector, this line might be contaminated by flux from other parts of Arp~299. No H$_{2}^{18}$O lines were detected (see Table \ref{tab:18oh}.)

\subsection{OH}
Five OH doublets, with lower energy levels up to ${\sim} 400$~K, were detected with PACS. The doublets were mainly seen in absorption, but the OH $^{2}\Pi_{1/2}-{^{2}\Pi}_{1/2}\, \frac{3}{2}-\frac{1}{2}$ doublet at $163$~$\mu$m was detected in emission. This doublet is often seen in emission, for example in NGC~4418 and Arp~220 \citepalias{gon12}, because it is usually excited by absorption in the the OH $^{2}\Pi_{1/2}-{^{2}\Pi}_{3/2}\, \frac{5}{2}-\frac{3}{2}$ and $^{2}\Pi_{1/2}-{^{2}\Pi}_{3/2}\, \frac{3}{2}-\frac{3}{2}$ doublets at $35$ and $53.3$~$\mu$m, respectively, followed by a cascade down to the ground state. Measured line parameters of the detected doublets are summarized in Table \ref{tab:pacslines} and the spectral ranges containing the lines are shown in Fig. \ref{fig:h2o_oh}. All lines are also indicated in the energy diagram of OH in Fig. \ref{fig:energy_diagram}. Most lines seem to be uncontaminated by other species, but the OH $^{2}\Pi_{3/2}-{^{2}\Pi}_{3/2}\, \frac{9}{2}^--\frac{7}{2}^+$ line at $65$~$\mu$m might be weakly contaminated by H$_{2}$O~$6_{25}\!\rightarrow\!5_{14}$. The extended line emission and absorption seen in the OH $^{2}\Pi_{3/2}-{^{2}\Pi}_{3/2}\, \frac{5}{2}-\frac{3}{2}$ doublet at $119$~$\mu$m combined with the strong redshift (${\sim}175$~km\,s$^{-1}$) in its absorption indicates the presence of an extended component with inflowing molecular gas. No $^{18}$OH doublets were detected (see Table \ref{tab:18oh}). We do not estimate an upper limit of the blue component of the $^{18}$OH $^{2}\Pi_{3/2}-{^{2}\Pi}_{3/2}\, \frac{5}{2}-\frac{3}{2}$ doublet at $120$~$\mu$m as it is close to an absorption in the CH$^{+}$ $3-2$ line. It is possible that some of the flux in this absorption comes from the blue component of the $^{18}$OH doublet, but we find this unlikely as the red component of this doublet is missing while the components of the $^{16}$OH $^{2}\Pi_{3/2}-{^{2}\Pi}_{3/2}\, \frac{5}{2}-\frac{3}{2}$ doublet at $119$~$\mu$m show almost equal strengths.

\begin{table*} 
  \caption{H$_{2}$O and OH lines detected with PACS}             
  \label{tab:pacslines}      
  \centering                          
  \begin{tabular}{l c c c c c c c c}       
    \hline\hline                 
    Line & $E_{\mathrm{lower}}$ & $\lambda_{\mathrm{rest}}$ & V$_{\mathrm{c}}\tablefootmark{a,b}$ & $\Delta V\tablefootmark{a,c}$ & $\Delta V_{\mathrm{inf}}\tablefootmark{d}$ & Cont.\tablefootmark{e} & Line peak\tablefootmark{a,f} & Flux\tablefootmark{a,f} \\  
    & (K) & ($\mu$m) & ($\mathrm{km\,\,s^{-1}}$) & ($\mathrm{km\,\,s^{-1}}$) & ($\mathrm{km\,\,s^{-1}}$) & (Jy) & (Jy) & ($\mathrm{Jy\,\,km\,\,s^{-1}} $) \\
    \hline                       
    H$_{2}$O $4_{22}\!\rightarrow\!3_{13}$ & $205$ & $57.636$ & $-7(12)$ & $346(30)$ & $280$ & $83.2$ & $-3.0(0.2)$ & $-1071(117)$  \\ 
    H$_{2}$O $5_{24}\!\rightarrow\!4_{13}$ & $396$ & $71.067$ & $34(25)$ & $283(62)$ & $229$ & $81.2$ & $-1.6(0.3)$ & $-476(133)$  \\ 
    H$_{2}$O $7_{17}\!\rightarrow\!6_{06}$ & $643$ & $71.540$ & $-4(24)$ & $147(59)$ & $\ldots$ & $82.3$ & $-0.9(0.3)$ & $-148(78)$  \\
    H$_{2}$O $7_{07}\!\rightarrow\!6_{16}$ & $644$ & $71.947$ & $-42(22)$ & $281(56)$ & $228$ & $83.0$ & $-1.3(0.2)$ & $-396(103)$  \\
    H$_{2}$O $3_{21}\!\rightarrow\!2_{12}$ & $114$ & $75.381$ & $36(2)$ & $320(6)$ & $279$ & $85.0$ & $-11.9(0.2)$ & $-4032(96)$  \\ 
    H$_{2}$O $6_{16}\!\rightarrow\!5_{05}$ & $468$ & $82.031$ & $15(16)$ & $358(39)$ & $330$ & $80.2$ & $-2.4(0.2)$ & $-900(126)$  \\
    H$_{2}$O $6_{06}\!\rightarrow\!5_{15}$ & $470$ & $83.284$ & $-8(17)$ & $272(42)$ & $235$ & $79.7$ & $-2.0(0.3)$ & $-578(115)$  \\
    H$_{2}$O $2_{21}\!\rightarrow\!1_{10}$ & $61$ & $108.073$ & $15(7)$ & $386(17)$ & $226$ & $65.5$ & $-5.7(0.2)$ & $-2348(133)$ \\
    H$_{2}$O $3_{13}\!\rightarrow\!2_{02}$ & $101$ & $138.528$ & $-33(21)$ & $446(53)$ & $356$ & $42.6$ & $-1.3(0.1)$ & $-600(93)$ \\
    H$_{2}$O $4_{13}\!\rightarrow\!4_{04}$ & $320$ & $187.111$ & $-53(22)$ & $385(55)$ & $343$ & $23.0$ & $1.1(0.2)$ & $469(87)$ \\
    OH    $^{2}\Pi_{3/2}-{^{2}\Pi}_{3/2}\, \frac{9}{2}^--\frac{7}{2}^+$ & $291$ &  $65.132$ & $13(5)$ & $387(13)$ & $341$ & $86.6$ & $-8.3(0.2)$ & $-3432(145)$  \\
    OH    $^{2}\Pi_{3/2}-{^{2}\Pi}_{3/2}\, \frac{9}{2}^+-\frac{7}{2}^-$ & $291$ & $65.279$ & $8(5)$ & $362(12)$ & $313$ & $86.6$ & $-8.6(0.2)$ & $-3331(138)$  \\ 
    OH    $^{2}\Pi_{1/2}-{^{2}\Pi}_{1/2}\, \frac{7}{2}-\frac{5}{2}$ & $416$ & $71.197\tablefootmark{g}$ & $31(8)$ & $405(22)$ & $\ldots$ & $81.4$ & $-5.4(0.2)$ & $-2311(159)$ \\ 
    OH    $^{2}\Pi_{3/2}-{^{2}\Pi}_{3/2}\, \frac{7}{2}^+-\frac{5}{2}^-$ & $121$ & $84.420$ & $23(3)$ & $321(8)$ & $292$  & $79.5$ & $-14.9(0.3)$ & $-5085(159)$ \\
    OH    $^{2}\Pi_{3/2}-{^{2}\Pi}_{3/2}\, \frac{7}{2}^--\frac{5}{2}^+$ & $121$ & $84.597$ & $39(3)$ & $309(8)$ & $278$ & $79.5$ & $-15.4(0.3)$ & $-5073(160)$  \\
    OH    $^{2}\Pi_{3/2}-{^{2}\Pi}_{3/2}\, \frac{5}{2}^--\frac{3}{2}^+$ & $0$ & $119.233$ & $172(8)$ & $241(18)$ & $\ldots$ & $73.9$ & $-7.3(0.5)$ & $-1871(187)$ \\
    OH    $^{2}\Pi_{3/2}-{^{2}\Pi}_{3/2}\, \frac{5}{2}^+-\frac{3}{2}^-$ & $0$ & $119.441$ & $172(8)$ & $238(20)$ & $\ldots$ & $73.9$ & $-6.8(0.5)$ & $-1722(187)$ \\
    OH    $^{2}\Pi_{1/2}-{^{2}\Pi}_{1/2}\, \frac{3}{2}^+-\frac{1}{2}^-$ & $182$ & $163.126$ & $11(11)$ & $413(27)$ & $343$ & $31.4$ & $4.9(0.2)$ & $2148(157)$ \\
    OH    $^{2}\Pi_{1/2}-{^{2}\Pi}_{1/2}\, \frac{3}{2}^--\frac{1}{2}^+$ & $182$ & $163.397$ & $33(10)$ & $417(23)$ & $347$ & $31.4$ & $5.3(0.2)$ & $2373(153)$ \\
    \hline                                   
  \end{tabular}
  \tablefoot{
    \tablefoottext{a}{Values from Gaussian fits to the lines, numbers in parenthesis indicate 1$\sigma$ uncertainties from these fits.}
    \tablefoottext{b}{Velocity shift of line center relative to $z = 0.010411$.}
    \tablefoottext{c}{FWHM of lines.}
    \tablefoottext{d}{Inferred velocity width is based on the instrument resolution assuming a Gaussian profile; this is not listed for doublets that are not well separated.}
    \tablefoottext{e}{Value of the fitted baseline at the line center.}
    \tablefoottext{f}{Absorption lines indicated with minus sign.}
    \tablefoottext{g}{The two $\Lambda-$components are (nearly) blended into a single spectral feature.}
  }
\end{table*}

\begin{table*} 
  \caption{$3\sigma$ upper limits to undetected H$_{2}^{18}$O and $^{18}$OH lines.}             
  \label{tab:18oh}      
  \centering                          
  \begin{tabular}{l c c c c}        
    \hline\hline                 
    Line & $\lambda_{\mathrm{rest}}$ & Line peak\tablefootmark{a} & Flux\tablefootmark{a,b} & Predicted flux\tablefootmark{c} \\  
    &($\mu$m) & (Jy) & ($\mathrm{Jy\,\,km\,\,s^{-1}} $) &  ($\mathrm{Jy\,\,km\,\,s^{-1}} $) \\
    \hline                        
    H$_{2}^{18}$O $4_{22}\!\rightarrow\!3_{13}$ & $57.840$ & $<0.99$ & $<210$ & $\ldots$  \\  
    H$_{2}^{18}$O $5_{24}\!\rightarrow\!4_{13}$ & $71.754$ & $<1.20$ & $<210$ & $\ldots$   \\
    H$_{2}^{18}$O $6_{06}\!\rightarrow\!5_{15}$ & $83.591$ & $<1.19$ & $<170$ & $\ldots$   \\
    H$_{2}^{18}$O $3_{13}\!\rightarrow\!2_{02}$ & $139.586$ & $<1.32$ & $<370$ & $\ldots$  \\
    H$_{2}^{18}$O $2_{20}\!\rightarrow\!2_{11}$ & $250.034$ & $<0.54$ & $<160$ & $\ldots$  \\  
    H$_{2}^{18}$O $4_{22}\!\rightarrow\!4_{13}$ & $252.167$ & $<0.51$ & $<150$ & $\ldots$   \\
    H$_{2}^{18}$O $3_{12}\!\rightarrow\!3_{03}$ & $253.762$ & $<0.48$ & $<140$ & $\ldots$   \\
    H$_{2}^{18}$O $3_{21}\!\rightarrow\!3_{12}$ & $263.738$ & $<0.51$ & $<160$  & $\ldots$  \\
    $^{18}$OH    $^{2}\Pi_{3/2}-{^{2}\Pi}_{3/2}\, \frac{9}{2}^--\frac{7}{2}^+$ & $65.543$ &  $<1.59$ & $<310$ & $-40$  \\
    $^{18}$OH    $^{2}\Pi_{3/2}-{^{2}\Pi}_{3/2}\, \frac{9}{2}^+-\frac{7}{2}^-$ & $65.690$ &  $<1.59$ & $<310$ & $-40$   \\ 
    $^{18}$OH    $^{2}\Pi_{1/2}-{^{2}\Pi}_{1/2}\, \frac{7}{2}-\frac{5}{2}$ & $71.679$ &  $<1.05$ & $<190$ & $-10$ \\  
    $^{18}$OH    $^{2}\Pi_{3/2}-{^{2}\Pi}_{3/2}\, \frac{7}{2}^+-\frac{5}{2}^-$ & $84.947$ &  $<1.62$ & $<230$ & $-230$ \\
    $^{18}$OH    $^{2}\Pi_{3/2}-{^{2}\Pi}_{3/2}\, \frac{7}{2}^--\frac{5}{2}^+$ & $85.123$ &  $<1.62$ & $<230$ & $-230$  \\
    $^{18}$OH    $^{2}\Pi_{3/2}-{^{2}\Pi}_{3/2}\, \frac{5}{2}^+-\frac{3}{2}^-$ & $120.171\tablefootmark{d}$ &  $<1.23$& $<380$ & $-330$  \\
    $^{18}$OH    $^{2}\Pi_{1/2}-{^{2}\Pi}_{1/2}\, \frac{3}{2}^+-\frac{1}{2}^-$ & $164.269$ &  $<0.90$& $<220$ & $30$  \\
    $^{18}$OH    $^{2}\Pi_{1/2}-{^{2}\Pi}_{1/2}\, \frac{3}{2}^--\frac{1}{2}^+$ & $164.545$&  $<0.90$ & $<220$ & $30$ \\
    \hline                                   
  \end{tabular}
  \tablefoot{
    \tablefoottext{a}{Upper limits are to the absolute values.}
    \tablefoottext{b}{Upper limits to the fluxes were calculated using the instrument resolution and assuming a Gaussian profile for lines observed with PACS and a sinc profile for lines observed with SPIRE.}
    \tablefoottext{c}{Predicted flux, for the $^{18}$OH lines, from the inner component using the limiting value $N_{\mathrm{OH}}/\tau_{50}=5\times10^{15}$, corresponding to a lower limit on the $^{16}$O/$^{18}$O ratio of $400$.}
    \tablefoottext{d}{The blue component of the $^{18}$OH doublet at $120$~$\mu$m is close to the CH$^{+}$ $3-2$ absorption line and we only estimate an upper limit for the red component.}
  }
\end{table*}

\begin{table*} 
  \caption{H$_{2}$O lines detected with SPIRE}             
  \label{tab:spirelines}      
  \centering                          
  \begin{tabular}{l c c c c c c c c}       
    \hline\hline                 
    Line & $E_{\mathrm{upper}}$ & $\lambda_{\mathrm{rest}}$ & V$_{\mathrm{c}}\tablefootmark{a,b}$ & $\Delta V\tablefootmark{a,c}$ & $\Delta V_{\mathrm{inf}}\tablefootmark{d}$ & Cont.\tablefootmark{e} & Line peak\tablefootmark{a} & Flux\tablefootmark{a} \\  

    & (K) &($\mu$m) & ($\mathrm{km\,\,s^{-1}}$) & ($\mathrm{km\,\,s^{-1}}$) & ($\mathrm{km\,\,s^{-1}}$) & (Jy) & (Jy) & ($\mathrm{Jy\,\,km\,\,s^{-1}} $) \\
    \hline                        
    H$_{2}$O $1_{11}\!\rightarrow\!0_{00}\tablefootmark{f}$ & $53$ & $269.272$ & $\ldots$ & $\ldots$ & $\ldots$ & $8.1$ & $<0.18$ & $<57$  \\
    H$_{2}$O $2_{02}\!\rightarrow\!1_{11}$ & $101$ & $303.456$ & $-24(12)$ & $382(36)$ & $\ldots$ & $5.7$ & $2.6(0.2)$ & $891(53)$ \\
    H$_{2}$O $2_{11}\!\rightarrow\!2_{02}\tablefootmark{g}$ & $137$ & $398.643$ & $-4(26)$ & $516(90)$ & $\ldots$ & $1.7$ & $1.7(0.3)$ & $809(80)$ \\
    H$_{2}$O $2_{20}\!\rightarrow\!2_{11}$ & $196$ & $243.974$ & $12(18)$ & $367(47)$ & $309(59)$ & $10.9$ & $1.8(0.2)$ & $652(156)$  \\
    H$_{2}$O $3_{12}\!\rightarrow\!3_{03}$ & $249$ & $273.193$ & $-25(15)$ & $457(41)$ & $415(44)$ & $7.8$ & $2.4(0.2)$ &$1132(153)$ \\
    H$_{2}$O $3_{21}\!\rightarrow\!3_{12}$ & $305$ & $257.795$ & $-11(8)$ & $375(21)$ & $304(28)$ & $9.3$ & $4.1(0.2)$ & $1528(178)$  \\
    H$_{2}$O $4_{22}\!\rightarrow\!4_{13}$ & $454$ & $248.247$ & $-4(20)$ & $389(55)$ & $322(65)$ & $10.4$ & $1.6(0.2)$ & $615(155)$  \\
    H$_{2}$O $5_{23}\!\rightarrow\!5_{14}$ & $642$ & $212.526$ & $-22(34)$ & $550(88)$ & $546(82)$ & $16.3$ & $1.1(0.2)$ & $628(125)$  \\
    \hline                                   
  \end{tabular}
  \tablefoot{
        \tablefoottext{a}{Values from fits to the spectral lines, numbers in parenthesis indicate 1$\sigma$ uncertainties from these fits.}
        \tablefoottext{b}{Velocity shift of line center relative to $z = 0.010411$.}
        \tablefoottext{c}{FWHM of lines.}
        \tablefoottext{d}{Inferred velocity width is based on the instrument resolution assuming a sinc profile, only listed for partially resolved lines.}
        \tablefoottext{e}{Value of the fitted baseline at the line center.}
        \tablefoottext{f}{Undetected, upper limit is $1\sigma$ value.}
        \tablefoottext{g}{Detected with the spectrometer long wavelength array, due to the larger beam size this line might be contaminated by flux from the other components of Arp~299.}
  }
\end{table*}


\section{Models}\label{sec:models}
We have modeled the observations using the spherically symmetric radiative transfer code described by \citet{gon97,gon99}, including a treatment of line overlaps for the OH $\Lambda$-components. Dust emission is included with the dust grains being simulated as a mixture of silicates and amorphous carbon, for which the adopted mass absorption coefficient as a function of wavelength is shown in \citet{gon14a}. The model parameters used to characterize the physical conditions in the source are the dust opacity at $100$~$\mu$m ($\tau_{100}$), the dust temperature ($T_{\mathrm{dust}}$), the gas temperature ($T_{\mathrm{gas}}$), the H$_{2}$ density ($n_{\mathrm{H_{2}}}$), and the column density of H$_{2}$O or OH per optical depth at $50$~$\mu$m ($N_{\mathrm{H_{2}O/OH}}/\tau_{50}$). Collisional rates with H$_{2}$ are taken from \citet{dub09} and \citet{dan11} for H$_{2}$O and from \citet{off94} for OH. Line broadening is simulated by a microturbulent velocity ($v_{\mathrm{tur}}$) which is set to $80$~km\,s$^{-1}$ in order to match the observed line widths, regardless of whether it is due to turbulence or velocity gradients along the line of sight. Our results on line ratios depend on $N_{\mathrm{H_{2}O/OH}}/\Delta V$. If rotation is (partially) responsible for line broadening, the velocity dispersion along each line of sight would be somewhat lower and thus $N_{\mathrm{H_{2}O/OH}}$ would be somewhat decreased. The column density ratios would be the same, but abundances would be somewhat lower. In all models we have adopted a gas-to-dust ratio of 100 by mass, similar to the average value in LIRGs reported by \citet{wil08}. We have assumed a covering factor of unity for all models.

\begin{figure}
  \centering
  \includegraphics[width=8.0cm]{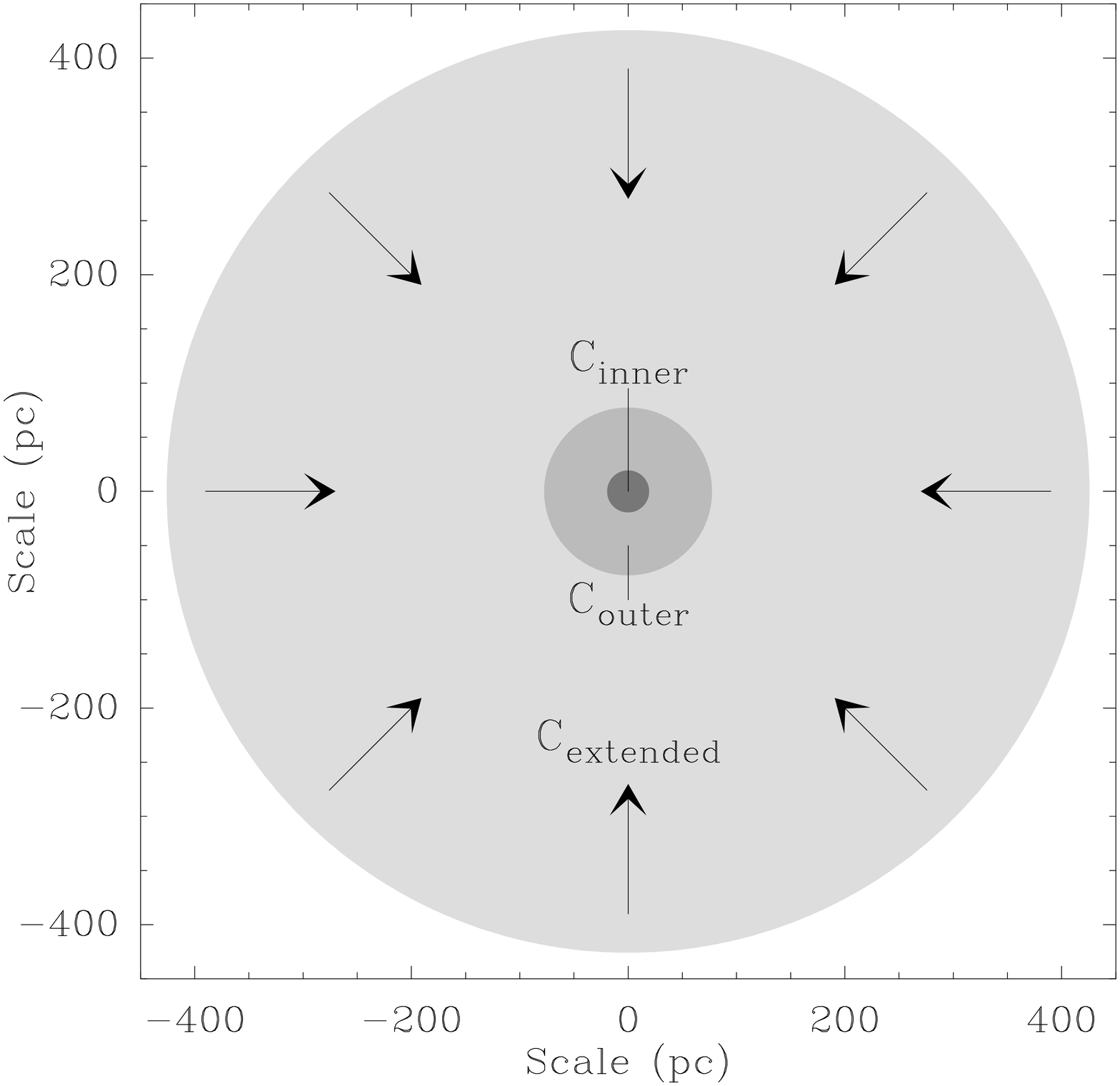}
  \caption{Schematic representation of the spherically symmetric model for the molecular components of Arp~299A showing approximate spatial scales. Most of the molecular absorption is formed in the outer layers of the dense, most compact C$_{\mathrm{inner}}$ component while the H$_{2}$O emission in the submillimeter seems to originate in the more optically thin C$_{\mathrm{outer}}$ component. A tentative inflowing C$_{\mathrm{extended}}$ component is included to account for the redshifted ground state OH doublet at $119$~$\mu$m.}
  \label{fig:model}
\end{figure}

The data presented in Sect. \ref{sec:observations} reveal a rich molecular line spectrum where the OH and H$_{2}$O lines in the far-IR PACS spectra are seen primarily in absorption while the H$_{2}$O lines in the submm SPIRE spectrum are seen in emission only. This division of absorption and emission lines between PACS and SPIRE is caused by the excitation of H$_{2}$O rotational levels through absorption in the far-IR lines, followed by de-excitation through emission in lines at longer (submm) wavelenghts where the optical depth of the dust is lower. Similar absorption/emission dichotomies are seen in Mrk~231 \citep{gon10} and Zw~049.057 \citep{fal15} and the excitation processes for the H$_{2}$O submm lines are further explored in \citet{gon14a}. The rich variety of emission and absorption lines cannot be reproduced by a single set of parameters as the high- and low-lying lines are expected to arise in different regions due to their differing excitation requirements. Similar to the inner regions of Arp~220, NGC~4418, and Zw~049.057 \citetext{\citetalias{gon12}; \citealp{fal15}} the high excitation seen in both H$_{2}$O and OH in Arp~299A cannot be accounted for by collisions alone. The dominant excitation mechanism is instead absorption of photons emitted by the warm dust in the central regions of the galaxy. In our modeling we try to reproduce the observed molecular absorption/emission using the least possible number of parameterized components and find that the lines are primarily formed in two components, which are discussed in Sects. \ref{sec:inner} and \ref{sec:outer} and summarized in Tables \ref{tab:cont} and \ref{tab:lines}. A tentative additional component which can account for the redshifted ground state OH $^{2}\Pi_{3/2}-{^{2}\Pi}_{3/2}\, \frac{5}{2}-\frac{3}{2}$ doublet at $119$~$\mu$m is discussed in Sect. \ref{sec:extended}. All three components are included in a schematic representation of the model in Fig. \ref{fig:model}. A short outline of the modeling is presented in Sect. \ref{sec:outline}.

\begin{table*} 
  \caption[]{Parameters of the continuum models.}
  \label{tab:cont}
  \centering
  \begin{tabular}{lcccccc}
    \hline\hline 
    \noalign{\smallskip}
    C\tablefootmark{a} & Radius\tablefootmark{b} & $T_{\mathrm{d}}$\tablefootmark{b} & $\tau_{100}$\tablefootmark{b} & $N_{\mathrm{H_{2}}}\tablefootmark{c}$ & $M\tablefootmark{d}$ & $L\tablefootmark{e}$ \\ 
    & (pc) & (K) &  & (cm$^{-2}$) & ($10^8$ M$_{\sun}$) & (L$_{\sun}$) \\
    \noalign{\smallskip}
    \hline
    $C_{\mathrm{inner}}$ & $20-25$ & $90-120$ & $1-4$ & $(1-3)\times10^{24}$ & $0.25-1.25$ & $(0.7-1.5)\times10^{11}$ \\
    $C_{\mathrm{outer}}$ & $50-100$  & $70-90$ &  $0.25-1$ & $(1.5-7)\times10^{23}$ & $0.25-5$ & $(1.7-3.0)\times10^{11}$ \\ 
    \noalign{\smallskip}
    \noalign{\smallskip}
    \hline
  \end{tabular} 
  \tablefoot{
    \tablefoottext{a}{Component.}
    \tablefoottext{b}{Independent parameter.}
    \tablefoottext{c}{Column density of H$_{2}$, calculated assuming a mass-absorption coefficient of 44.4 cm$^2$~g$^{-1}$ at 100 $\mu$m and a gas-to-dust mass ratio of 100.}
    \tablefoottext{d}{Estimated gas mass, assuming spherical symmetry.}
    \tablefoottext{e}{Unattenuated dust luminosity of the component.}
  }
\end{table*}

\begin{table*} 
  \caption[]{Derived molecular column densities and abundances.}
  \label{tab:lines}
  \centering
  \begin{tabular}{lcccccc}
    \hline\hline 
    \noalign{\smallskip}
    C\tablefootmark{a} & $N_{\mathrm{H_2O}}/\tau_{50}\tablefootmark{b,c}$ & $N_{\mathrm{OH}}/\tau_{50}\tablefootmark{b,c}$ &   $N_{\mathrm{^{18}OH}}/\tau_{50} \tablefootmark{b,c}$ & $\chi_{\mathrm{H_2O}}\tablefootmark{d}$ & $\chi_{\mathrm{OH}}\tablefootmark{d}$\\
    &  (cm$^{-2}$)  &  (cm$^{-2}$) & (cm$^{-2}$) & & \\
    \noalign{\smallskip}
    \hline
    $C_{\mathrm{inner}}$  & $(0.5-2)\times10^{18}$ & $(0.2-1)\times10^{19}$  &  <$5\times10^{15}$  & $(1-5)\times10^{-6}$ & $(0.5-5)\times10^{-5}$ \\
    $C_{\mathrm{outer}}$ & $(0.2-1)\times10^{17}$ & $(2-8)\times10^{17}$ &  <$5\times10^{15}$  & $(0.5-2.5)\times10^{-7}$  & $(0.5-1)\times10^{-6}$ \\
    \noalign{\smallskip}
    \noalign{\smallskip}
    \hline
  \end{tabular}
  \tablefoot{
    \tablefoottext{a}{Component.}
    \tablefoottext{b}{Independent parameter.}
    \tablefoottext{c}{Column density per unit of dust opacity at 50~$\mu$m, $\tau_{50}$.}
    \tablefoottext{d}{Estimated molecular abundance relative to H nuclei, with an estimated column of H nuclei per $\tau_{50}$ of $4\times10^{23}$~cm$^{-2}$.}
  }
\end{table*}

\subsection{Outline of the modeling}\label{sec:outline}
Our general approach to the modeling was to compare the observed ratios of various H$_{2}$O lines to a grid of models and then estimate the radius ($R$) from the scaling to the total flux of the lines. As the high-lying ($E_{\mathrm {lower}}\gtrsim 300$~K) lines are expected to be uncontaminated by absorption from a lower-excitation component they were fitted first. To account for the missing flux in the low-lying lines the procedure was then repeated for them, but now taking the contribution from the first component into account. Finally, the OH lines were reproduced by fitting an OH column to the models developed for H$_{2}$O. The details of the modeling in the two components are discussed in Sects. \ref{sec:inner} and \ref{sec:outer}. An approach where the inner and outer model components were fitted simultaneously was also tried, and yielded similar results.

\subsection{Inner component}\label{sec:inner}
The high-lying ($E_{\mathrm {lower}}\gtrsim 300$~K) lines of H$_{2}$O provide the best constraints for the parameters in the inner regions of the galaxy. We have compared the observed fluxes of the five highest-lying lines detected with PACS and the two highest-lying lines detected with SPIRE (see Figs. \ref{fig:h2o_oh}-\ref{fig:h2o_sled}), all normalized to the flux of the H$_{2}$O $58$~$\mu$m $4_{22}\!\rightarrow\!3_{13}$ line, to a grid of models where we have varied the parameters $\tau_{100}$, $T_{\mathrm{dust}}$, $T_{\mathrm{gas}}$, $n_{\mathrm{H_{2}}}$, and $N_{\mathrm{H_{2}O}}/\tau_{50}$. In each grid point we have computed the reduced $\chi^{2}$ for the line ratios:
\begin{equation}\label{eq:sumofsquares}
  \chi^{2}=\frac{1}{N-n}\sum\limits_{i=1}^{N}\frac{(\mathrm{obs}_{i}-\mathrm{model}_{i})^{2}}{\sigma_{i}^{2}},
\end{equation}
were $N$ is the number of line ratios fitted, $n$ is the number of fitted parameters, $\mathrm{obs}_{i}$ and $\mathrm{model}_{i}$ are the observed and modeled values of line ratio $i$, and $\sigma_{i}$ is the standard deviation of line ratio $i$, calculated through error propagation. Parameter ranges are determined based on where the $\chi^{2}$ has increased by at least $20$\% from its minimum value.

\begin{figure}
  \centering
  \includegraphics[width=8.0cm]{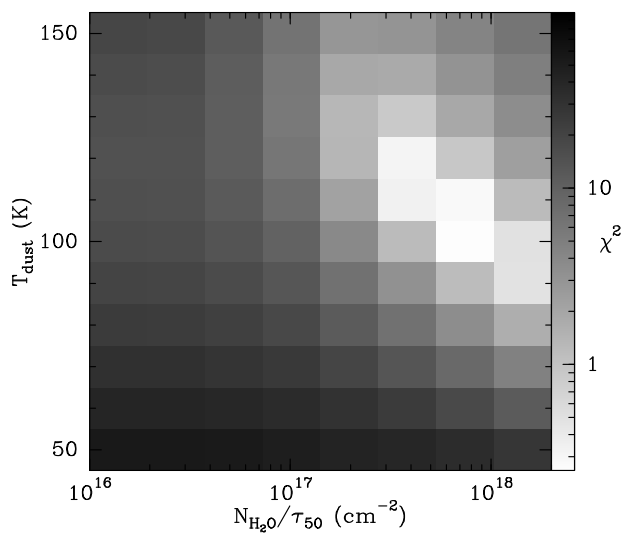}
  \caption{Reduced $\chi^{2}$ for the observed high-lying H$_{2}$O lines in our grid of models with $\tau_{100}=2$.}
  \label{fig:h2o_modelgrid_high}
\end{figure}

We find that the best fit to the observations is achieved with a dust opacity of $\tau_{100}=1-4$, with $\tau_{100}=2$ being the preferred value. Because of the high values of the Einstein coefficients of the OH and H$_{2}$O lines, for values of $n_{\mathrm{H_{2}}}\leq3\times10^{6}$~cm$^{-3}$ collisions have negligible impact on the results. In Fig. \ref{fig:h2o_modelgrid_high} we show the reduced $\chi^{2}$ in our grid of models for the preferred dust opacity of $\tau_{100}=2$. The relative fluxes of the high-lying lines can be well fitted with different combinations of dust temperatures between $T_{\mathrm{dust}}=90$ and $120$~K, and H$_{2}$O columns between $N_{\mathrm{H_{2}O}}/\tau_{50}=5\times10^{17}$ and $2\times10^{18}$~cm$^{-2}$, with lower $T_{\mathrm{dust}}$ corresponding to higher $N_{\mathrm{H_{2}O}}/\tau_{50}$. To account for the absolute flux in the lines a radius of the component between $R=20$ and $25$~pc is required. All parameter ranges are summarized in Tables \ref{tab:cont} and \ref{tab:lines}. The best fit, which is also included in Figs. \ref{fig:continuum}-\ref{fig:h2o_sled}, is achieved with $\tau_{100}=2$, $T_{\mathrm{dust}}=100$~K, $N_{\mathrm{H_{2}O}}/\tau_{50}=1\times10^{18}$~cm$^{-2}$ and $R=22$~pc. This component accounts for both the high-lying absorption lines in the far-IR and the high-lying ($E_{\mathrm{upper}}>400$~K) emission lines in the submillimeter. The column density of H$_{2}$ is calculated from the value of $\tau_{100}$ assuming a mass-absorption coefficient of 44.4 cm$^2$~g$^{-1}$ at 100 $\mu$m and a gas-to-dust mass ratio of 100. The molecular mass of the component is then calculated from this value, assuming spherical symmetry. Finally, the dust luminosity is based on the radius, temperature, and opacity of the component.

\begin{figure}
  \centering
  \includegraphics[width=8.0cm]{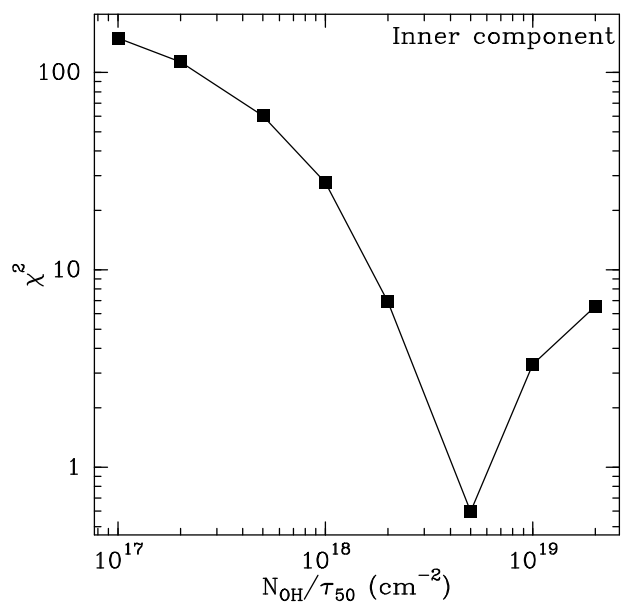}
  \caption{$\chi^{2}$ for the OH $^{2}\Pi_{1/2}-{^{2}\Pi}_{1/2}\, \frac{7}{2}-\frac{5}{2}$ at $71$~$\mu$m with different values of $N_{\mathrm{OH}}/\tau_{50}$ applied to the best fit model of the inner component found using the H$_{2}$O lines.}
  \label{fig:oh_models_high}
\end{figure}

OH $^{2}\Pi_{1/2}-{^{2}\Pi}_{1/2}\, \frac{7}{2}-\frac{5}{2}$ at $71$~$\mu$m, the transition with the highest lower-state energy, is the only doublet that is expected to be formed in the inner component only. The $\chi^{2}$ for the OH $71$~$\mu$m $^{2}\Pi_{1/2}-{^{2}\Pi}_{1/2}\, \frac{7}{2}-\frac{5}{2}$ is presented in Fig. \ref{fig:oh_models_high} for different values of $N_{\mathrm{OH}}/\tau_{50}$ applied to the best fit model found for H$_{2}$O. We find that the flux is best reproduced in the range $N_{\mathrm{OH}}/\tau_{50}=2\times10^{18}$ to $1\times10^{19}$~cm$^{-2}$, corresponding to an OH/H$_{2}$O ratio of $2-10$.

An upper limit  to the $^{18}$OH column density was estimated using the $3\sigma$ flux upper limits of the three lowest-lying doublets $^{18}$OH $^{2}\Pi_{3/2}-{^{2}\Pi}_{3/2}\, \frac{7}{2}-\frac{5}{2}$, $^{2}\Pi_{3/2}-{^{2}\Pi}_{3/2}\, \frac{5}{2}-\frac{3}{2}$, and $^{2}\Pi_{1/2}-{^{2}\Pi}_{1/2}\, \frac{3}{2}-\frac{1}{2}$ at $85$, $120$, and $164$~$\mu$m, respectively. The non-detections of these doublets are not compatible with an $^{18}$OH column density higher than $N_{\mathrm{^{18}OH}}/\tau_{50}=5\times10^{15}$~cm$^{-2}$, corresponding to a lower limit of $400$ to the $^{16}$OH/$^{18}$OH ratio. Predicted flux values using this limiting value are included in Table \ref{tab:18oh}.

\subsection{Outer component}\label{sec:outer}
\begin{figure}
  \centering
  \includegraphics[width=8.0cm]{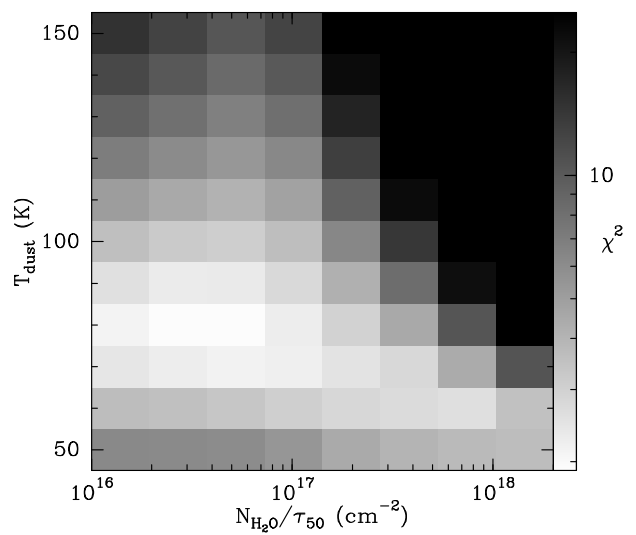}
  \caption{Reduced $\chi^{2}$ for the observed low-lying H$_{2}$O lines in our grid of models with $\tau_{100}=0.5$ for the outer component, taking the contribution from the inner component into account.}
  \label{fig:h2o_modelgrid_low}
\end{figure}

Even in the best fit model for the inner component there is flux missing in the low-lying OH and and H$_{2}$O lines detected with PACS and, especially, the low-lying H$_{2}$O lines detected with SPIRE. To account for this missing flux in the models, a less excited and more extended component is required. To constrain the parameters of this outer component we have  followed the same general procedure as for the inner component, but now for the low-lying lines normalized to the flux of the H$_{2}$O $303$~$\mu$m $2_{02}\!\rightarrow\!1_{11}$ line. In this process we have also taken the best fit model for the inner component into account. The observed line ratios are best reproduced with a dust opacity in the range $\tau_{100}=0.25-1$. In this component collisional excitation becomes non-negligible for $n_{\mathrm{H_{2}}}>1\times10^{5}$~cm$^{-3}$, but in these models the undetected H$_{2}$O $269$~$\mu$m $1_{11}\!\rightarrow\!0_{00}$ line goes into emission, favoring models with lower $n_{\mathrm{H_{2}}}$ where collisional excitation is negligible. The reduced $\chi^{2}$ for our grid of models at the preferred dust opacity $\tau_{100}=0.5$ is presented in Fig. \ref{fig:h2o_modelgrid_low}. The line ratios are best reproduced with dust temperatures between $T_{\mathrm{dust}}=70$ and $90$~K, and H$_{2}$O columns between $N_{\mathrm{H_{2}O}}/\tau_{50}=2\times10^{16}$ and $1\times10^{17}$~cm$^{-2}$. To also account for the absolute fluxes, the component should have a radius of $R=50-100$~pc. The best fit is achieved with $\tau_{100}=0.5$, $T_{\mathrm{dust}}=70$~K, $N_{\mathrm{H_{2}O}}/\tau_{50}=5\times10^{16}$~cm$^{-2}$ and $R=75$~pc. 

\begin{figure}
  \centering
  \includegraphics[width=8.0cm]{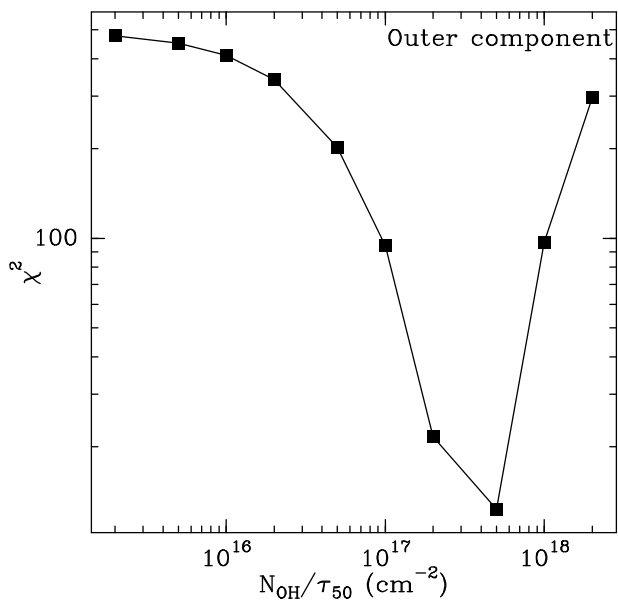}
  \caption{Reduced $\chi^{2}$ for the OH doublets, except for OH $119$~$\mu$m $^{2}\Pi_{3/2}-{^{2}\Pi}_{3/2}\, \frac{5}{2}-\frac{3}{2}$, with different values of $N_{\mathrm{OH}}/\tau_{50}$ applied to the best fit model of the outer component found using the H$_{2}$O lines.}
  \label{fig:oh_models_low}
\end{figure}

In Fig. \ref{fig:oh_models_low} we present the reduced $\chi^{2}$ for the OH doublets in the outer component for different values of $N_{\mathrm{OH}}/\tau_{50}$ applied to the best fit model found for H$_{2}$O, now taking the OH contribution from the inner component into account. The OH $119$~$\mu$m $^{2}\Pi_{3/2}-{^{2}\Pi}_{3/2}\, \frac{5}{2}-\frac{3}{2}$ doublet which is redshifted by ${\sim}175$~km\,s$^{-1}$ is not included here. The best fit is achieved with a column density of OH in the range $N_{\mathrm{OH}}/\tau_{50}=2\times10^{17}$ to $8\times10^{17}$~cm$^{-2}$, corresponding to an OH/H$_{2}$O ratio of $4-16$. Applying a similar process to the $3\sigma$ upper limits of the $^{18}$OH doublets we get an upper limit to the $^{18}$OH column density of $N_{\mathrm{^{18}OH}}/\tau_{50}<5\times10^{15}$~cm$^{-2}$.

\subsection{Extended component}\label{sec:extended}
Most of the flux in both the H$_{2}$O and the OH lines is reproduced with the two components described in Sects. \ref{sec:inner} and \ref{sec:outer}. The ground state OH $^{2}\Pi_{3/2}-{^{2}\Pi}_{3/2}\, \frac{5}{2}-\frac{3}{2}$ doublet at $119$~$\mu$m is however redshifted by ${\sim}175$~km\,s$^{-1}$ compared with the models, and the continuum at wavelengths $\gtrsim50$~$\mu$m is underpredicted by the combined central components. One way to shift the ground state OH doublet and simultaneously reproduce some of the continuum at longer wavelengths is to include a cooler and more extended component with a uniform inward velocity field. In this component the OH will mostly be in the ground state and produce redshifted absorption from the gas in front of the nucleus as well as emission at systemic velocity from gas on the sides, effectively shifting the ground state OH doublet to longer wavelengths while leaving the other transitions mostly unchanged.

\begin{figure}
  \centering
  \includegraphics[width=8.0cm]{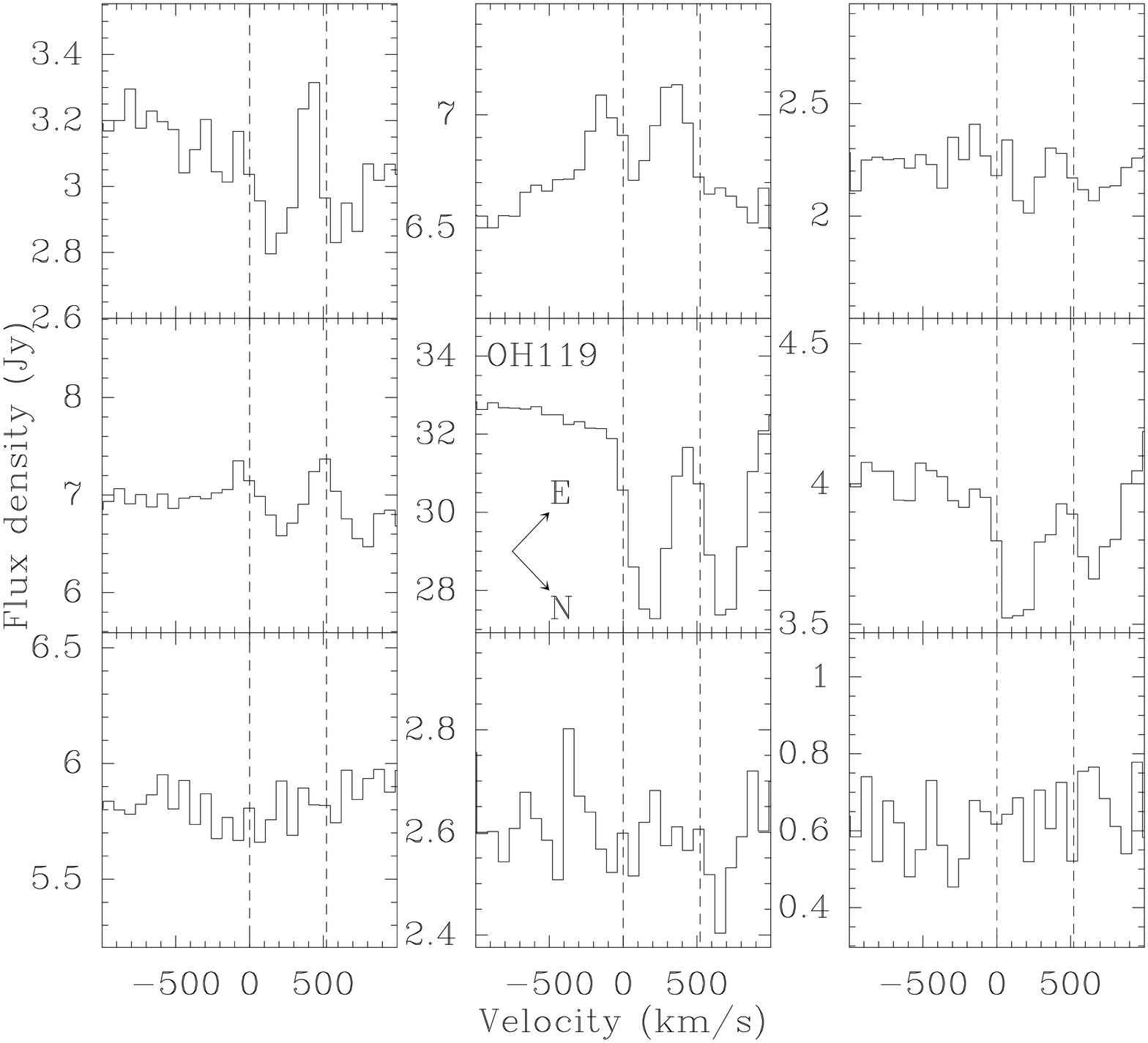}
  \caption{Central nine spaxels in the observation of the ground state OH $^{2}\Pi_{3/2}-{^{2}\Pi}_{3/2}\, \frac{5}{2}-\frac{3}{2}$ doublet at $119$~$\mu$m. The compass in the central spaxel shows the approximate orientation of the array on the sky. We note that, where detected, the absorption features are redshifted, and the emission features are blueshifted relative to the systemic velocity. Toward the southern spaxel (upper-left panel), both redshifted absorption and blueshifted emission are observed, i.e., an inverse P-Cygni profile.}
  \label{fig:oh119_3x3}
\end{figure}

Our limited angular resolution precludes an accurate quantitative analysis of the inflow energetics, but the lack of OH $119$~$\mu$m $^{2}\Pi_{3/2}-{^{2}\Pi}_{3/2}\, \frac{5}{2}-\frac{3}{2}$ absorption at systemic velocities, where the OH columns are enormous, suggests some cancellation between the expected absorption and emission at blueshifted velocities. This blueshifted emission is indeed seen in neighboring spaxels (Fig. \ref{fig:oh119_3x3}), and the combined redshifted absorption and surrounding blueshifted emission is consistent with an extended inflow. The geometry of the inflowing gas may significantly depart from sphericity, and thus the values we give, based on our spherical radiative transfer approach, should be considered a first approach; further high-angular observations will be required to better constrain the energetics.

In Fig. \ref{fig:continuum} as well as in the OH panels of Fig. \ref{fig:h2o_oh} we have included an extended component with parameters $R=400$~pc, $\tau_{100}=0.25$, $T_{\mathrm{dust}}=400$, $N_{\mathrm{OH}}/\tau_{50}=4\times10^{15}$~cm$^{-2}$ and a uniform velocity for the OH of $200$~km\,s$^{-1}$ towards the nucleus. Based on the total OH column and assuming a hemispherical inflow with the same OH abundance as in the outer component, a rough estimate of the mass inflow rate is $14-28$~M$_{\sun}$\,yr$^{-1}$. With only one doublet available we do not attempt a full analysis of this component and the parameters are not well constrained, but the mass inflow rate we obtain is comparable to that inferred in NGC~4418 \citetext{\citetalias{gon12}; \citealp{cos13,sak13}}. We note that, in Fig. \ref{fig:continuum}, the continuum of the combined components still significantly falls short of the observed data. Part of this discrepancy is likely the result of contamination from the other two components of the system, but there is probably also regions of even more extended continuum emission, not associated with a molecular component.

\subsection{Underpredicted lines}\label{sec:underpredicted}
Even with three components included, not all data is well represented by our best-fit models. In the SPIRE range, the H$_{2}$O $2_{11}\!\rightarrow\!2_{02}$ line at $399$~$\mu$m (see last panel of Fig. \ref{fig:h2o_spire}) is underpredicted by our models. As noted in Sect. \ref{sec:datareduction} this is the only SPIRE line that is detected with just the spectrometer long wavelength array, and due to the larger beam size it might be contaminated by flux from the other components of Arp 299. As a comparison, the H$_{2}$O $2_{02}\!\rightarrow\!1_{11}$ line at $303$~$\mu$m, which is situated in the overlap region of the two SPIRE detectors, shows ${\sim}30\%$ more flux in the long wavelength array than in the short wavelength array. 

  In the PACS range, the H$_{2}$O $4_{13}\!\rightarrow\!4_{04}$ line at $187$~$\mu$m (see middle panel of last row in Fig. \ref{fig:h2o_oh}) is also clearly underpredicted by our models. A possible explanation is that the wavelength of this line places it in a transition region between the absorption lines at shorter wavelengths, where the optical depth is higher, and the emission lines at longer wavelengths, where the optical depth is lower. It is likely that such a line cannot be well predicted by our simple models which use distinct, uniform components.

\section{Discussion}\label{sec:discussion}
\subsection{The gas and dust components of Arp~299A}
The properties of the two main components responsible for the molecular spectrum, described in Sect. \ref{sec:models}, are summarized in Tables \ref{tab:cont} and \ref{tab:lines}. The components have sizes comparable to those of giant molecular clouds (GMCs) in the Galaxy but both are warmer as well as significantly more massive and luminous than typical GMCs, which have masses and luminosities on the order of $10^{5}-10^{6}$~M$_{\sun}$ and $10^{5}-10^{8}$~L$_{\sun}$, respectively \citep[e.g][]{sco89}.

With an estimated column density of $N_{\mathrm{H_{2}}}{\sim}10^{24}$~cm$^{-2}$ in the inner component, Arp~299A seems to be somewhat less obscured than the CONs in Arp~220, NGC~4418 and Zw~049.057 \citetext{\citetalias{gon12}; \citealp{fal15}}, but still close to the limit to be considered Compton-thick ($N_{\mathrm{H}}\gtrsim1.5\times10^{24}$~cm$^{-2}$). Compared to these other CONs, the inner component of Arp~299A has many similarities with the warm component in NGC~4418 \citepalias{gon12} and also to the core component of Zw~049.057 \citep{fal15}. The outer component of Arp~299A is more similar to the extended component of NGC~4418 \citepalias{gon12} or to the outer component of Zw~049.057 \citep{fal15}, which however is estimated to be cooler than that of Arp~299A.

As in Zw~049.057 and NGC~4418, the central regions of Arp~299A are extreme compared to the center of the Milky Way, where a comparable amount of molecular gas ($\mathrm{few}\times10^{7}$~M$_{\odot}$) is distributed on much larger ($200-500$~pc) scales \citep[e.g.][]{cox89,mor96,fer07}. Our results show that also galaxies with luminous [C {\sc ii}] may have hidden CONs, and thus have nuclei with very young, compact, and embedded activity.

\subsubsection{Inflowing gas in the nucleus?}\label{sec:inflow}
The presence of gas flowing towards the nucleus is suggested by the redshift of ${\sim}175$~km\,s$^{-1}$ in the OH ground state $^{2}\Pi_{3/2}-{^{2}\Pi}_{3/2}\, \frac{5}{2}-\frac{3}{2}$ doublet at $119$~$\mu$m. An alternative explanation to this feature could be that it is formed in low-excited gas in the foreground, not directly associated with the central regions of Arp~299A. However, similar, but weaker and \emph{less redshifted}, features are seen in some of the neighboring PACS spaxels along with weak blueshifted emission, indicating that the feature is indeed associated with the nucleus. As the gas seems to be inflowing on spatial scales larger than the two central components, it is not as highly excited and we only see it in the ground state OH transition. Based on our rudimentary modeling of the redshifted doublet (see Sect. \ref{sec:extended}), assuming a hemispherical inflow, we estimate an upper limit to the mass inflow rate of $14-28$~M$_{\sun}$\,yr$^{-1}$. This molecular inflow could provide the nucleus with fresh material.

\subsubsection{Molecular abundances}
We find high abundances, relative to H nuclei, of H$_{2}$O and OH in the inner component, $1-5\times10^{-6}$ and $0.5-5\times10^{-5}$, respectively. The H$_{2}$O and OH abundances found in the outer component are $0.5-2.5\times10^{-7}$ and $0.5-1\times10^{-6}$, respectively. We thus see high OH/H$_{2}$O ratios of $\gtrsim1$ in both components.

The H$_{2}$O abundances in the inner component are comparable to those found in the eastern nucleus of Arp~220 \citepalias{gon12} as well as in Zw~049.057 \citep{fal15}. In the Galaxy, there are no regions of comparable size with H$_{2}$O abundances this high, but on smaller scales hot cores \citep[e.g. in NGC~6334~I and Orion~KL;][]{emp13,mel10} can have abundances as high as $10^{-5}$. In the outer component the H$_{2}$O abundances are similar to those found in the warm component of NGC~4418 \citepalias{gon12}. The OH abundances in the inner component are more similar to those in the western nucleus of Arp~220 and the core component of NGC~4418 \citepalias{gon12}. In the Galaxy, a similar OH abundance was found in the shocked low-velocity gas around the compact infrared source Orion-IRc2 by \citet{wri00}.

To account for the high H$_{2}$O and OH abundances we consider three main scenarios: ion-neutral reactions initialized by cosmic-ray or X-ray ionization of H$_{2}$ \citep{van13}, sublimation of H$_{2}$O ice from dust grains when the dust temperature rises above ${\sim}100$~K \citep{fra01,van13}, and neutral-neutral reactions in warm gas ($T\gtrsim250$~K) \citep{neu95,van13}. If the high abundances are due to sublimation from dust-grains, they will eventually decrease unless the core is replenished with icy dust grains by the inflow or there is another formation process to efficiently counter H$_{2}$O and OH destruction by, for example, photodissociation. In the other two scenarios, the abundances can persist as long as the formation conditions stay the same. 

In the inner component, where the highest abundances are found, the dust temperature is $90-120$~K, enough for sublimation to occur. As we are unable to put any constraints on the gas temperature, we cannot rule out neutral-neutral reactions in warm gas. In the ion-neutral scenario, H$_{2}$O and OH are formed through dissociative recombination of H$_{3}$O$^{+}$ with a branching ratio that favors OH \citep{her90,jen00}. Although we do not exclude the possibility of contributions from the other two processes, the fact that we see high (${>}1$) OH/H$_{2}$O abundance ratios in both components hints that ion-neutral reactions play an important part. Models of dense molecular gas irradiated by high rates of cosmic-rays or X-rays predict OH/H$_{2}$O ratios that can go above unity, with abundances that are consistent with our results \citep{mal96,mei05,mei11}.

We note that the highest-lying OH doublets, OH $^{2}\Pi_{1/2}-{^{2}\Pi}_{1/2}\, \frac{7}{2}-\frac{5}{2}$ and $^{2}\Pi_{3/2}-{^{2}\Pi}_{3/2}\, \frac{9}{2}-\frac{7}{2}$ at $71$ and $65$~$\mu$m, in Arp~299A have higher strengths, relative to lower-lying OH doublets and high-lying H$_{2}$O lines, than in Arp~220, Zw~049.057, or even NGC~4418 \citetext{\citetalias{gon12}; \citealp{fal15}}. This explains the very high abundance we infer for OH, compared to the other galaxies.

\subsection{Evolutionary stage - the $^{16}$O/$^{18}$O ratio}\label{sec:evolution}
The secondary nuclide $^{18}$O is produced in massive stars and it is expected to increase in relation to the primary nuclide $^{16}$O with each new generation of stars \citep{pra96}. The abundance ratio of $^{16}$O to $^{18}$O can therefore reflect the initial mass function (IMF) as well as the amount of stellar processing that the gas has gone through so far. A potential problem is that, under certain conditions, the small difference in molecular binding energies might cause the oxygen isotopes to be unequally distributed among molecules. \citet{lan84} found that such chemical fractionation effects for oxygen isotopes are small under all conditions, and our results for OH can be used directly to estimate the $^{16}$O/$^{18}$O ratio. 

\citet{gon12} use the $^{16}$O/$^{18}$O ratio as part of a tentative evolutionary scenario in which Mrk~231 with its extreme $^{16}$O/$^{18}$O ratio of ${\gtrsim}30$ \citep{fis10,gon14} has a more evolved starburst than Arp~220 \citepalias{gon12} and NGC~4418 which have ratios of ${\sim}70$ and ${\sim}500$, respectively \citepalias{gon12}. In this scenario Zw~049.057, with a ratio of $50-100$ \citep{fal15}, can be placed at a similar evolutionary stage as Arp~220. This of course assumes that the starbursts in these sources are of comparable sizes in relation to the host galaxies, that they have similar IMF and mixing of the ISM, and that the models are correct.

We do not detect any transitions of the $^{18}$O isotopologues of OH or H$_{2}$O in Arp~299A, but from the upper limits of the observed $^{18}$OH lines we estimate a lower limit of the $^{16}$O/$^{18}$O ratio of $400$ in the inner component. In the outer component we find a similar upper limit of the $^{18}$OH column, but due to the lower estimated $^{16}$OH column in this component, we only get a lower limit of the $^{16}$O/$^{18}$O ratio of $40$. In the case of the undetected H$_{2}^{18}$O lines, the corresponding H$_{2}^{16}$O lines are not strong and an upper limit of the H$_{2}^{18}$O column would be close to our estimate for the H$_{2}^{16}$O column. A lower limit of the $^{16}$O/$^{18}$O ratio estimated from the undetected H$_{2}^{18}$O transitions would therefore be low and not very meaningful. While the inner component with its ratio of $>400$, reminiscent of NGC~4418, shows no signs of strong stellar processing, we cannot rule out the presence of an evolved starburst in the outer component, which has a lower limit that would put it close to Mrk~231 on the evolutionary sequence.

\subsubsection{What governs the $^{16}$O/$^{18}$O ratio?}
Another interesting similarity between the ``young'' NGC~4418 and Arp~299A is the fact that they both show signs of molecular gas inflowing onto the nucleus but lack outflow signatures. NGC~4418 exhibits redshifted absorption by ${\sim}100$~km\,s$^{-1}$ in the [O {\sc i}] 63~$\mu$m line and the ground state OH $^{2}\Pi_{3/2}-{^{2}\Pi}_{3/2}\, \frac{5}{2}-\frac{3}{2}$ doublet at $119$~$\mu$m \citepalias{gon12} while the same ground state doublet is redshifted by $175$~km\,s$^{-1}$ in Arp~299A (see Sect. \ref{sec:inflow}). Together with the higher $^{16}$O/$^{18}$O ratio, this might indicate that the nuclei of both galaxies are in an earlier stage of evolution, before the onset of mechanical feedback. If this is the case, the ratio can be used as a tool to estimate the age of any starburst activity as it should decrease with every new generation of stars. Of course, this will also depend on how much of the gas reservoir is used by the starburst and how much remains unused. The high ratio could also simply mean that the molecular gas reservoirs in Arp~299A and NGC~4418 are replenished with relatively unprocessed gas, thus keeping the $^{16}$O/$^{18}$O ratio higher than in galaxies without inflows of molecular gas. A similar scenario seems to apply for the merger NGC~1614 \citep{kon16}.

\subsection{The nuclear power source}
The surface brightness of the inner component is high, ${\sim}5\times10^{13}$~L$_{\odot}$\,kpc$^{-2}$ over $~50$~pc, but its Compton-thick nature makes it difficult to unambiguously identify the embedded power source. Although this surface brightness is high for dusty radiation-pressure supported starbursts \citep{tho05} under normal conditions, it is still well below the theoretical value of ${\sim}10^{15}$~L$_{\odot}$\,kpc$^{-2}$ that can be attained in hot ($T_{\mathrm{d}}>200$~K) starbursts \citep{and11}. Such dust temperatures are not seen in our models, but due to the high obscuration we cannot rule out their existence.

\citet{fis14} used the \emph{Cloudy} \citep{fer13} spectral synthesis code to create one-dimensional models of a gas and dust cloud, centrally illuminated by either a starburst or an AGN. They explored hydrogen column densities, $N_{\mathrm{H}}$, up to $10^{25}$~cm$^{-2}$ and ionization parameters, $U$, between $10^{-4}$ and $1$ for hydrogen densities at the illuminated face of the cloud, $n_{\mathrm{H^{+}}}$, of $30$, $300$ and $3000$~cm$^{-3}$. In these idealized models, a central starburst with $n_{\mathrm{H^{+}}}=30$~cm$^{-3}$ can in principle reproduce the high OH column densities and high OH/H$_{2}$O ratios that we find in Arp~299A. The higher density starburst models, however, underpredict both the column density of OH and the OH/H$_{2}$O ratio. The AGN models on the other hand are able to produce higher OH column densities and OH/H$_{2}$O ratios for all three values of $n_{\mathrm{H^{+}}}$, resulting in better agreement with our results in Arp~299A for a wide range of parameters. These high rates of OH/H$_{2}$O are due to X-rays due to the AGN. Another possibility, not included in their starburst models, is that high rates of cosmic-ray ionization drive the chemistry. We note that these models assume a single, centrally concentrated, radiation source. Thus, we cannot rule out a distributed starburst or a combination of an AGN and a starburst.

Based on very long baseline interferometry (VLBI) radio observations, \citet{per10} concluded that there is a low-luminosity AGN in the central regions of Arp~299A. The bolometric luminosity of this putative AGN was estimated by \citet{alo13} to be ${\sim}1.6\times10^{10}$~L$_{\odot}$, corresponding to $10-20\%$ of the IR luminosity of the inner component in our models and thus not enough to be the dominant power source. The presence of a nuclear starburst in Arp~299A is supported by the many supernovae and supernova remnants detected in the central ${\sim}100$~pc using VLBI \citep[e.g.][]{nef04,ulv09,per09}. \citet{her12} analyzed the radial distribution of supernovae in the nuclear starburst of Arp~299A and found that it is consistent with an exponential disk with a scale length of $20-30$~pc around the low-luminosity AGN reported by \citet{per10}. The spatial scale of this star formation indicates that it might be associated with the inner component of our models.  

\subsection{Is there an outflow?}\label{sec:outflow}
Signatures of molecular outflows have been observed in other CON galaxies \citetext{e.g. Arp~220, NGC1377, and Zw~049.057; \citetalias{gon12}; \citealp{aal12,fal15}}. We do not see an outflow in Arp~299A, but a possible signature is the blueshifted H$_{2}$O maser which \citet{tar11} associate with the expanding structure first suggested by \citet{baa90}. In NGC~3256, which, like Arp~299A, is a merger LIRG, \citet{sak14} detected an outflow with a maximum velocity of $>750$~km\,s$^{-1}$ and a mass outflow rate of $>60$~M$_{\odot}$\,yr$^{-1}$. Furthermore, OH observations have revealed massive molecular outflows in a number of ULIRGs \citep[e.g.][]{fis10,stu11} and in systematic searches \citet{vei13} and \citet{spo13} found evidence of high velocity outflows in about two thirds of their (U)LIRG/QSO and ULIRG samples, respectively. We note that the processes responsible for outflows in less luminous galaxies may be very different from those of the most powerful outflows, which are found in ULIRGs and galaxies hosting AGNs \citep[see also][]{cic14}. So why do we not see an outflow in Arp~299A?

With a luminosity of ${\sim}1.6\times10^{10}$~L$_{\odot}$, estimated by \citep{alo13}, and assuming a velocity of $500$~km\,s$^{-1}$, radiation pressure from the AGN should be able to drive an outflow of ${\sim}0.6$~M$_{\odot}$\,yr$^{-1}$. \citet{bon12} estimated a lower limit to the core-collapse supernova rate in Arp~299A of $\nu_{\mathrm{SN}}>0.8$~yr$^{-1}$. Using this lower limit in Eqs. (10) and (34) in \citet{mur05}, and assuming that $10$\% of the energy is transferred to the ISM, we find that the supernovae could potentially drive a $500$~km\,s$^{-1}$ outflow of $32-50$~M$_{\odot}$\,yr$^{-1}$. Assuming an age of $10$~Myr, the total outflow mass would then be $6\times10^{6}-5\times10^{8}$~M$_{\odot}$. If outflowing directly towards us, an outflow of this magnitude should be clearly detectable in absorption towards the continuum source. However, if oriented differently, projection effects or obscuration from the galaxy could make an outflow hard to detect. For example, with the assumed velocity of $500$~km\,s$^{-1}$ and a $45$\degr\ inclination, a given parcel of outflowing gas could occult the ${<}100$~pc outer component for up to ${\sim}3\times10^{5}$~yr, reducing the possible amount of outflowing gas in front of the continuum source to $1.8\times10^{5}-1.5\times10^{7}$~M$_{\odot}$.

\subsection{Outlook}\label{sec:outlook}
The surface brightness of the inner component is consistent with either an AGN, a nuclear starburst, or a combination of the two, as the source(s) of its luminosity. An obscured AGN could also explain the high OH/H$_{2}$O ratios seen in both the inner and outer components, with the luminosity estimated by \citet{alo13} (${\sim}1.6\times10^{10}$~L$_{\odot}$) it would account for up to $20$\% of the IR luminosity from the inner component. As there is also evidence for starburst activity at the scale of our inner component \citep{her12}, we deem it most likely that a composite AGN/starburst is responsible for the high luminosity emanating from the central regions of Arp~299A. This nuclear activity could be fed by the infalling gas, provided that there is some process present to transport the gas all the way down to the central tens of parsecs of the nucleus. If this is the case, the replenishment of unprocessed gas might also explain the fact that the $^{16}$O/$^{18}$O ratio is higher than in similar galaxies, despite the presence of a nuclear starburst. Further study of this inflowing component and how it relates to the inner regions of Arp~299A could shed some light on how the nuclear gas concentrations in CONs are assembled and how their activity is sustained.  


\section{Conclusions}\label{sec:conclusions}
The spectroscopic observations of Arp~299A presented in this paper gave the following basic results:
\begin{itemize}
   \item High excitation in H$_{2}$O and OH is revealed with PACS spectroscopy. A total of nine H$_{2}$O lines and four OH doublets with lower level energies up to $E_{\mathrm{lower}} \sim 600$~K and $E_{\mathrm{lower}} \sim 400$~K, respectively, were detected in absorption. In addition, one H$_{2}$O line and one OH doublet were detected in emission.
   \item The ground state OH $^{2}\Pi_{3/2}-{^{2}\Pi}_{3/2}\, \frac{5}{2}-\frac{3}{2}$ doublet at $119$~$\mu$m is redshifted by ${\sim}175$~km\,s$^{-1}$ and is the only line with significant emission outside of the central spaxel of PACS, indicating an inflow of low-excited molecular gas. 
   \item SPIRE spectroscopy shows seven submillimeter H$_{2}$O emission lines with upper state energies up to $E_{\mathrm{upper}} \sim 650$~K. The ground state H$_{2}$O $269$~$\mu$m $1_{11}\!\rightarrow\!0_{00}$ line is however not detected. 
   \item No $^{18}$OH or H$_{2}^{18}$O lines were detected.
\end{itemize}

The observed lines in the absorption dominated PACS spectra and the emission dominated SPIRE spectrum as well as the continuum levels have been analyzed using multicomponent radiative transfer modeling. From this analysis we draw the following conclusions:
\begin{itemize}
   \item A high H$_{2}$ column density ($(1-3)\times 10^{24}$~cm$^{-2}$) towards the inner component indicates that it is Compton-thick. The dust temperature in this component is estimated to be $T_{\mathrm{d}}=90-120$~K. 
   \item High OH column densities per unit of continuum optical depth at $50$~$\mu$m of $(0.2-2)10^{19}$~cm$^{-2}$ and a high OH/H$_{2}$O ratio $>1$ indicate that ion-neutral chemistry induced by X-rays or cosmic-rays is important in the nucleus of Arp~299A.
   \item A high surface brightness, ${\sim}5\times 10^{13}$~L$_{\sun}$~kpc$^{-2}$ on a scale of ${\sim}50$~pc, in the inner component indicates that the luminosity is powered by a buried AGN and/or a nuclear starburst. Our analysis indicates that a composite source is the most likely.
   \item The non-detection of $^{18}$OH results in a $^{16}$O/$^{18}$O ratio of $>400$ in the core of Arp~299A. This is similar to the ratio found in NGC~4418, and might indicate either that the starbursts of Arp~299A and NGC~4418 are in an early evolutionary stage or that their gas reservoirs are replenished with relatively unprocessed gas through molecular inflows.

\end{itemize}

\begin{acknowledgements}
We thank the anonymous referee for a thorough and constructive report that helped improve the paper.
  NF and SA thank the Swedish National Space Board for generous grant support (grant numbers 145/11:1B, 285/12 and 145/11:1-3).
  E.G-A is a Research Associate at the Harvard-Smithsonian Center for Astrophysics, and thanks the Spanish Ministerio de Econom\'{\i}a y Competitividad for support under project FIS2012-39162-C06-01 and  ESP2015-65597-C4-1-R, and NASA grant ADAP NNX15AE56G.
  Basic research in IR astronomy at NRL is funded by the US-ONR; J.F. acknowledges support from NHSC/ JPL subcontracts 1435724 and 1456609.
  PACS has been developed by a consortium of institutes led by MPE (Germany) and including UVIE (Austria); KU Leuven, CSL, IMEC (Belgium); CEA, LAM (France); MPIA (Germany); INAFIFSI/OAA/OAP/OAT, LENS, SISSA (Italy); IAC (Spain). This development has been supported by the funding agencies BMVIT (Austria), ESA-PRODEX (Belgium), CEA/CNES (France), DLR (Germany), ASI/INAF (Italy), and CICYT/MCYT (Spain).
SPIRE has been developed by a consortium of institutes led by Cardiff University (UK) and including Univ. Lethbridge (Canada); NAOC (China); CEA, LAM (France); IFSI, Univ. Padua (Italy); IAC (Spain); Stockholm Observatory (Sweden); Imperial College London, RAL, UCL-MSSL, UKATC, Univ. Sussex (UK); and Caltech, JPL, NHSC, Univ. Colorado (USA). This development has been supported by national funding agencies: CSA (Canada); NAOC (China); CEA, CNES, CNRS (France); ASI (Italy); MCINN (Spain); SNSB (Sweden); STFC, UKSA (UK); and NASA (USA).
  This research has made use of NASA's Astrophysics Data System (ADS) and of GILDAS software (http://www.iram.fr/IRAMFR/GILDAS).

\end{acknowledgements}

\bibliographystyle{bibtex/aa}
\bibliography{ref}

\end{document}